\RequirePackage{fix-cm}
\documentclass[twocolumn,epjc3]{svjour3}  
\smartqed  

\RequirePackage{graphicx}
\RequirePackage{dcolumn}
\RequirePackage{bm}
\RequirePackage{color} 
\RequirePackage{xcolor} 
\RequirePackage{color} 
\RequirePackage{graphics}
\RequirePackage{float}
\RequirePackage{bm}        
\RequirePackage{amssymb}   
\RequirePackage{slashed}   
\RequirePackage{amsmath}   
\RequirePackage{verbatim}  
\RequirePackage{subfig}
\RequirePackage{mathptmx}      
\RequirePackage{flushend}
\RequirePackage[numbers,sort&compress]{natbib}
\RequirePackage[colorlinks,citecolor=blue,urlcolor=blue,linkcolor=blue]{hyperref}

\newcounter{YJC}

\journalname{Eur. Phys. J. C}

\begin{document}
\title{Optimize the event selection strategy to study the anomalous quartic gauge couplings at muon colliders using the support vector machine and quantum support vector machine}

\author{Shuai Zhang\thanksref{e1,addr1}
        \and
        Yu-Chen Guo\thanksref{e2,addr1}
        \and
        Ji-Chong Yang\thanksref{cr,e3,addr1,addr2}
}

\institute{Department of Physics, Liaoning Normal University, Dalian 116029, China\label{addr1} 
           \and
          Center for Theoretical and Experimental High Energy Physics, Liaoning Normal University, Dalian 116029, China\label{addr2}}

\thankstext[$\star$]{cr}{Corresponding author}
\thankstext{e1}{e-mail: 2802368240@qq.com}
\thankstext{e2}{e-mail: ycguo@lnnu.edu.cn}
\thankstext{e3}{e-mail: yangjichong@lnnu.edu.cn}

\date{Received: date / Revised version: date}

\maketitle

\begin{abstract}
The search of the new physics~(NP) beyond the Standard Model is one of the most important topics in current high energy physics. 
With the increasing luminosities at the colliders, the search for NP signals requires the analysis of more and more data, and the efficiency in data processing becomes particularly important.
As a machine learning algorithm, support vector machine~(SVM) is expected to to be useful in the search of NP.
Meanwhile, the quantum computing has the potential to offer huge advantages when dealing with large amounts of data, which suggests that quantum SVM~(QSVM) is a potential tool in future phenomenological studies of the NP.
How to use SVM and QSVM to optimize event selection strategies to search for NP signals are studied in this paper.
Taking the tri-photon process at a muon collider as an example, it can be shown that the event selection strategies optimized by the SVM and QSVM are effective in the search of the dimension-8 operators contributing to the anomalous quartic gauge couplings.
\end{abstract}

\maketitle

\section{\label{sec1}Introduction}

At present, the Standard Model~(SM) has been remarkably successful. 
However, it still faces many unsolved mysteries and limitations.
The SM cannot explain phenomena such as dark matter, the unification of gravity with quantum mechanics, etc. 
The existence of new physics~(NP) beyond the SM can help explain these mysteries. 
The search of NP has become one of the frontiers in high energy physics~(HEP)~\cite{Ellis:2012zz}. 
It can be expected that, future colliders will have large luminosities, which will not only increase the chances of detecting NP signals but also will require higher efficiency in detecting NP signals.

One way to explore NP efficiently is to use the SM effective field theory~(SMEFT) to explore possible NP signals~\cite{Weinberg:1979sa,Grzadkowski:2010es,Willenbrock:2014bja,Mass__2014}.
In SMEFT, NP effects are manifested in high dimensional operators with Wilson coefficients suppressed by powers of an NP scale $\Lambda$. 
By considering only the NP signals most likely to be observed, only a finite number of operators need be considered.
It is efficient in the sense that one can search for or reject many different NP models by examining one operator contributed by the many NP models.
The SMEFT focuses primarily on dimension 6 operators, but from a phenomenological point of view there are many cases where dimension 6 operators are absent but dimension 8 operators appear~\cite{Born:1934gh,Ellis:2018cos,Ellis:2017edi,Ellis:2017edi,Ellis:2019zex,Ellis:2020ljj,Gounaris:2000dn,Gounaris:1999kf,Senol:2018cks,Fu:2021mub,Degrande_2014,Jahedi:2022duc,Jahedi:2023myu}. 
And the dimension 8 operators are important from the perspective of convex geometry on the operator space~\cite{positivity1,positivity2,positivity3}. 
For one generation of fermions, there are $895$ baryon number conserving dimension-8 operators~\cite{Henning:2015alf,Anders:2018oin}, and kinematic analysis has to be done for each operator. 
The efficiency decreases as the number of operators to be considered increases.

In recent years, machine learning algorithms~(ML) have experienced rapid development in the field of HEP~\cite{Radovic:2018dip,Baldi:2014kfa,Ren:2017ymm,Abdughani:2018wrw,PhysRevLett.124.010508,Ren:2019xhp,DAgnolo:2018cun,Yang:2022fhw,Yang:2021ukg,Jiang:2021ytz,MdAli:2020yzb,Fol:2020tva}, including the use of deep learning architectures such as convolutional neural networks~\cite{Berthold:2023dys,Begue:2023ers}, transformers~\cite{Wu:2024thh,Usman:2024hxz,Nikolskaiia:2024ory}, and graph neural networks~\cite{Correia:2024ogc,Caillou:2024dcn, Vijayakumar:2024myb,Mitra:2024few} for various tasks.
From an ML perspective, the search for NP signals can be seen as anomaly detection~(AD)~\cite{DeSimone:2018efk,MdAli:2020yzb,Fol:2020tva,Kasieczka:2021xcg,Guo:2021jdn,Yang:2021kyy,Dong:2023nir,CrispimRomao:2020ucc,vanBeekveld:2020txa,Kuusela:2011aa,Zhang:2023khv}. 
Discriminating between NP signal events and background events can also be viewed as a classification problem, and there are a large number of classification algorithms available in ML. 
In this paper, we will explore a commonly used classification algorithm, the support vector machine~(SVM)~\cite{Boser1992ATA}.

It is widely accepted that quantum computing can be used to speed up computation. 
Researchers have already explored the application of SVM on quantum computers, leading to the development of quantum SVM~(QSVM)~\cite{Rebentrost_2014,Havlicek:2018nqz,Schuld:2019bfg}.
QSVM takes advantage of quantum computing, such as quantum superposition and quantum entanglement, to speed up the solution of SVM problems~\cite{Ozpolat:2023zwq,Akter:2023iwy,Aksoy:2023okt}.
The QSVM has already been used to analyze the $H\to \mu^+\mu^-$ process and the $\bar{t}tH$ production at the large hadron collider~(LHC)~\cite{Wu:2020cye,Wu:2021xsj}, and the process $e^+e^-\to ZH$ at the circular electron-positron collider~(CEPC)~\cite{Fadol:2022umw}.
While SVM is well suited for data classification, its potential for NP search remains uncertain, due to the fact that algorithms for classification problems are not always suitable for AD, the latter being concerned with situations where it is equivalent to categorizing two classes that are very different in number.
In this paper, we focus on investigating the feasibility of using SVM and QSVM to search for NP.
One of the main advantages of using ML is that the whole procedure is fully automated.
Following this spirit, it will be shown in this paper that, given a space of observables, the SVM can find an event selection strategy automatically.

As an example, in this paper, we study dimension-8 operators contributing to anomalous quartic gauge couplings~(aQGCs).
Since the vector boson scattering processes are sensitive to the aQGCs, the LHC has maintained a close interest in aQGCs~\cite{ATLAS:2014jzl,CMS:2020gfh,ATLAS:2017vqm,CMS:2017rin,CMS:2020ioi,CMS:2016gct,CMS:2017zmo,CMS:2018ccg,ATLAS:2018mxa,CMS:2019uys,CMS:2016rtz,CMS:2017fhs,CMS:2019qfk,CMS:2020ypo,CMS:2020fqz}.
Phenomenological studies on aQGCs are also being conducted intensively~\cite{Green:2016trm,Chang:2013aya,Anders:2018oin,Zhang:2018shp,Bi:2019phv,Guo:2020lim,Guo:2019agy,Yang:2021pcf}.
Recently, there has been increasing interest in muon colliders, as they offer the potential to reach high energies and high luminosities while providing a cleaner experimental environment with less impact from QCD backgrounds than the hadron colliders~\cite{Buttazzo_2018,delahaye2019muon,Costantini:2020stv,Lu:2020dkx,AlAli:2021let,Franceschini:2021aqd,Palmer_1996,Holmes:2012aei,Liu:2021jyc,Liu:2021akf}.
It is expected that the future muon colliders have large luminosities. 
As a testing ground, we take the muon colliders to study the aQGCs, the event selection strategy and expected coefficient constraints are studied using the SVM and QSVM.
We consider the tri-photon process at a muon collider which is known to be particularly sensitive to the transverse operators contributing to the aQGCs~\cite{Yang:2020rjt}, and is therefore a process worth investigating.
Note that, as an ML algorithm, the procedure to optimize the event selection strategy using the SVM and QSVM does not depend on the process under study.

The remainder of this paper is structured as follows. 
Section~\ref{sec2} is a brief introduction to the aQGCs and the tri-photon process.
Section~\ref{sec3} discusses the event selection strategy using the SVM.
The numerical results of the SVM are presented in section~\ref{sec4}.
The implementation of the QSVM and the numerical results are presented in section~\ref{sec5}.
Section~\ref{sec6} compares the expected coefficient constraints of several algorithms. 
Section~\ref{sec7} summarizes the main conclusions.

\section{\label{sec2}aQGCs and tri-photon process at the muon colliders}

\begin{table}[hbtp]
\centering
\begin{tabular}{c|c|c|c}
\hline
coefficient & constraint & coefficient & constraint \\
\hline
$f_{T_{0}}/\Lambda^4$ &$[-0.12, 0.11]$~\cite{CMS:2019qfk}&$f_{T_{6}}/\Lambda^4$&$[-0.4, 0.4]$~\cite{CMS:2020ypo} \\
$f_{T_{1}}/\Lambda^4$ &$[-0.12, 0.13]$~\cite{CMS:2019qfk}&$f_{T_{7}}/\Lambda^4$&$[-0.9, 0.9]$~\cite{CMS:2020ypo} \\
$f_{T_{2}}/\Lambda^4$ &$[-0.28, 0.28]$~\cite{CMS:2019qfk}&$f_{T_{8}}/\Lambda^4$&$[-0.43, 0.43]$~\cite{CMS:2020fqz} \\
$f_{T_{5}}/\Lambda^4$ &$[-0.5, 0.5]$~\cite{CMS:2020ypo}&$f_{T_{9}}/\Lambda^4$&$[-0.92, 0.92]$~\cite{CMS:2020fqz}\\
\hline
\end{tabular}
\caption{The expected constraints on the $O_{T_{i}}$ coefficients obtained at $95~\%$ C.L. at the LHC.}
\label{table:1}
\end{table}

The dimension-8 operators contributing to the aQGCs can be classified as scalar/longitudinal operators $O_{S_i}$, mixed transverse and longitudinal operators $O_{M_i}$ and transverse operators $O_{T_i}$~\cite{Alessandro_2018}.
In this paper we focus on $O_{T_i}$ operators~\cite{Eboli:2006wa,Eboli:2016kko}, which contribute to tri-photon process at muon colliders. 
They are,
\begin{equation}
\begin{split}
&O_{T,0}={\rm Tr}\left[\widehat{W}_{\mu\nu}\widehat{W}^{\mu\nu}\right]\times {\rm Tr}\left[\widehat{W}_{\alpha\beta}\widehat{W}^{\alpha\beta}\right],\\
&O_{T,1}={\rm Tr}\left[\widehat{W}_{\alpha\nu}\widehat{W}^{\mu\beta}\right]\times {\rm Tr}\left[\widehat{W}_{\mu\beta}\widehat{W}^{\alpha\nu}\right],\\
&O_{T,2}={\rm Tr}\left[\widehat{W}_{\alpha\mu}\widehat{W}^{\mu\beta}\right]\times {\rm Tr}\left[\widehat{W}_{\beta\nu}\widehat{W}^{\nu\alpha}\right],\\
&O_{T,5}={\rm Tr}\left[\widehat{W}_{\mu\nu}\widehat{W}^{\mu\nu}\right]\times B_{\alpha\beta}B^{\alpha\beta},\\
&O_{T,6}={\rm Tr}\left[\widehat{W}_{\alpha\nu}\widehat{W}^{\mu\beta}\right]\times B_{\mu\beta}B^{\alpha\nu},\\
&O_{T,7}={\rm Tr}\left[\widehat{W}_{\alpha\mu}\widehat{W}^{\mu\beta}\right]\times B_{\beta\nu}B^{\nu\alpha},\\
&O_{T,8}=B_{\mu\nu}B^{\mu\nu}\times B_{\alpha\beta}B^{\alpha\beta},\\
&O_{T,9}=B_{\alpha\mu}B^{\mu\beta}\times B_{\beta\nu}B^{\nu\alpha},\\
\end{split}
\label{eq.2.1}
\end{equation}
where $\widehat{W}\equiv \vec{\sigma}\cdot {\vec W}/2$ with $\sigma$ being the Pauli matrices and ${\vec W}=\{W^1,W^2,W^3\}$, $B_{\mu}$ and $W_{\mu}^i$ are $U(1)_{\rm Y}$ and $SU(2)_{\rm I}$ gauge fields, and $B_{\mu\nu}$ and $W_{\mu\nu}$ correspond to the field strength tensors.
Table~\ref{table:1} presents the constraints obtained at the LHC.

\begin{figure}[htbp]
\begin{center}
\includegraphics[width=0.8\hsize]{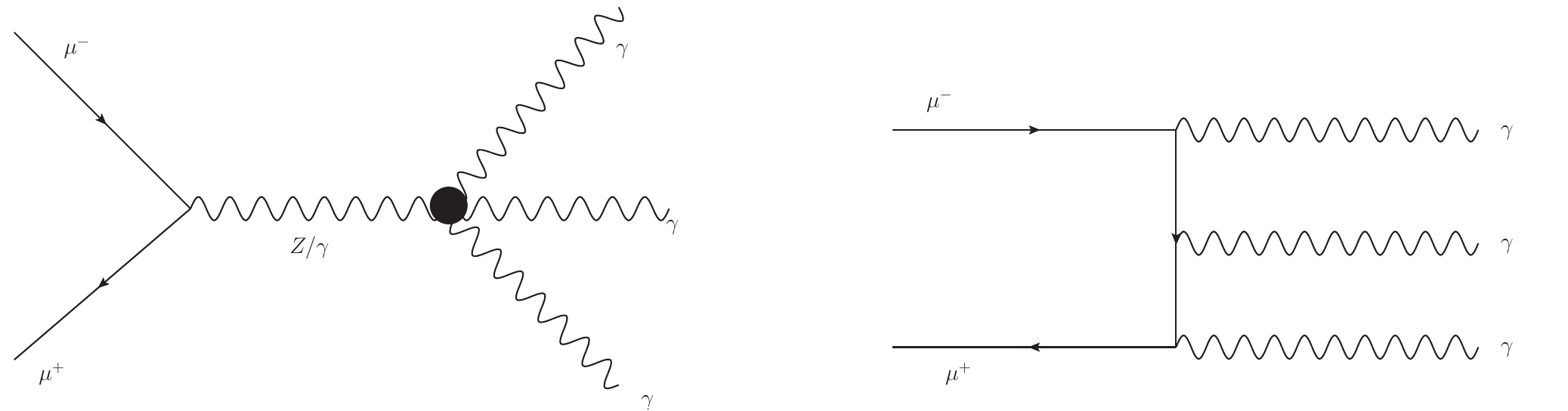}
\caption{\label{fig:Feyman} Feynman diagrams for the tri-photon process $\mu^{+}\mu^{-}\rightarrow\gamma\gamma\gamma$ . 
The left panel shows the diagrams induced by $O_{T_i}$, while the right panel is one of the diagrams in the SM. 
There are other five diagrams in the SM which can be obtained by permuting the photons in the final state.}
\end{center}
\end{figure}

For the process $\mu^+\mu^-\to \gamma\gamma\gamma$, the diagrams induced by the $O_{T_i}$ operators are shown in the left panel of Fig.~\ref{fig:Feyman}, and the right panel of Fig.~\ref{fig:Feyman} shows one of the diagrams in the SM. 
In the SM there are five other diagrams that can be obtained by permuting the photons in the final state.
The expressions for the contributions of the $O_{T_i}$ operators and the interference between the $O_{T_i}$ operators and the SM are given in Ref.~\cite{Yang:2020rjt}, where it is shown that the contributions of the $O_{T_{1,6}}$ operators are exactly the same as those of the $O_{T_{0,5}}$ operators.
Therefore, in the following we will concentrate on $O_{T_{0,2,5,7,8,9}}$ operators.

\section{\label{sec3}Optimize the event selection strategy using SVM}

The search of the NP signals at the colliders with high luminosities is to look for a small number of anomalies in a vast amount of background data.
Ignoring the interference, this is also a binary classification problem that distinguish the NP events from the SM background.
One of the ML algorithms designed for the binary classification problem is the SVM. 
Therefore, it can be expected that the SVM can be used to search for NP signals.

A collision event can be represented as a point in a multidimensional space, where each dimension corresponds to an observable in the measurement~(denoted as $M_i$) associated with that collision event, which is also known in ML as the `feature space'.
Traditional event selection strategies typically use intervals for segmentation, i.e. retaining or excluding events with $M_i^{\rm min}<M_i<M_i^{\rm max}$. 
This is equivalent to selecting or excluding a hyper cuboid in feature space. 
Unlike the traditional approach, if we use the SVM algorithm for the event selection strategy, we will automatically obtain a hyperplane that separates the NP from the SM in the feature space. 
This hyperplane corresponds to a criterion consisting of $M_i$, so similar to the traditional event selection strategy, it is also a strategy consisting of observables and has the form of an inequality, i.e. selecting events with $f(M_i)>0$, where $f(x)$ is some function. 
Since the SVM can automatically find the hyperplane that best separates the two types of events, the event selection strategy obtained by the SVM can be viewed as an optimization of a traditional one.
Although some AD algorithms, such as principal component analysis AD~\cite{Dong:2023nir} and isolation forest~\cite{Guo:2021jdn}, have clear geometric meanings, it is still difficult for them to form an explicit event selection strategy. 
An explicit event selection strategy has a clearer physical meaning, is interpretable, and reflects the laws of NP. 
Therefore, it can be used not only to increase signal significance, but also to provide guidance on what features of the NP signal to look for.
It should be emphasized that the collection of observables does not require any knowledge of the NP model, but only a sufficiently large number of observables based on the final state to be investigated.

Another similar approach to automatically optimize the event selection strategy is to use the decision tree, which is also an algorithm used for classification~\cite{Roe:2004na,Friedman:2003ic,Breiman:1984jka}.
With continuous improvement of the algorithms, decision tree and algorithms derived from decision trees are widely used in HEP, including the search of NP signals~\cite{Hanson:2018uhf,alhroob2012search,Wen:2014mha}. 
The use of SVM is similar, but with a different algorithm.
Since there is no a priori superiority of any classification algorithm over the others, the best algorithm for a particular task is often task-dependent.
The study in this paper will enrich the toolbox for the phenomenological study of NP, in particular providing a tool that can be implemented in quantum computing.

\subsection{\label{sec3.1}A brief introduction to soft margin SVM}

Since our data set is not linearly divisible, the soft margin SVM is used in this paper.
It has been shown that finding the best hyperplane is equivalent to solving an optimization problem, which can be expressed as~\cite{Cortes:1995hrp},
\begin{equation}
\begin{aligned} 
\underset{\vec{w},b,\xi _i }{\min}\ \frac{1}{2}|\vec{w}|^2+c\sum_{i=1}^N{\xi _i}, \\
 s.t.\ \ \ \ y_i(w^{j}x_i^{j}+b)\ge 1-\xi _i,\;(i = 1,2,...N),\\
 \xi _i\ge 0,\ (i=1,2,...,N), \\
\end{aligned} 
\label{eq.three}
\end{equation}
where $\vec{w}$ is the normal vector of the hyperplane, which can determine the direction of the hyperplane and $w^{j}$ is the $j$-th component of $\vec{w}$; $|\vec{w}|$ denotes the length of $\vec{w}$, $|\vec{w}|^2=\sum _j(w^{j})^2$. 
$b$ is the bias term, which can determine the distance between the hyperplane and the origin.
$c$ denotes the penalty parameter, which is a hyper-parameter that can be adjusted on its own, and usually takes the value of $10$. 
$\xi _{i}$ is the relaxation factor, which can be expressed as $\xi _i=\max \left( 0,1-y_i\left( \sum _j w^{j}x_{i}^{j}+b \right) \right)$.
$x_{i}^{j}$ is the $j$-th feature of the $i$-th data point, $y_{i}$ is the class label, assigned as $+1$ and $-1$ to represent two different classes~($+1$ for the NP and $-1$ for the SM in our case).
$N$ is the number of data points. 
Of all the variables listed above, $\vec{x}_i$ and $y_i$ are known, $c$ is an adjustable parameter, and $\vec{w}$ and $b$ are the ones to be solved for.
After $w^{j}$ and $b$ are obtained, and the hyperplane equation can be expressed as,
\begin{equation}
\begin{split} 
 \sum_{j}w^{j}x_{i}^{j}+b=0.
\end{split} 
\end{equation}
\label{eq.w} 

\subsection{\label{sec3.2}Data preparation}

\begin{table}[htbp]
\centering
\begin{tabular}{c|c|c|c|c|c} 
\hline
  & $ p^{1}$& $ p^{2}$ & $p^{3}$ & $p^{4}$& $p^{5}$ \\
 \hline
 observable & $E_{1}$& $E_{2}$&$E_{3}$&$p_{T_1}$&$p_{T_2}$ \\
  \hline
  & $ p^{6}$ & $ p^{7} $ & $p^{8} $ & $ p^{9}$& $p^{10}$ \\  
  \hline
  observable & $p_{T_3}$& $\eta_{1}$&$\eta_{2}$ & $\eta_{3}$ & $\Delta R_{12}$ \\
  \hline
  & $ p^{11}$& $p^{12}$ & $ p^{13}$ & $ p^{14}$& $p^{15}$\\  
  \hline
  observable & $\Delta R_{13}$& $\Delta R_{23}$&$m_{12}$&$m_{13}$&$m_{23}$\\
  \hline  
\end{tabular}
\caption{In this paper, the events are mapped into points in a 15-dimensional space where the axes are observables, and a point is denoted as $\vec{p}$. The components of $\vec{p}$ and the corresponding observables are listed.}
\label{table:12}
\end{table}

In this paper, the events are generated using Monte Carlo~(MC) simulation with the help of \verb"MadGraph5@NLO" toolkit~\cite{Alwall:2014hca,Christensen:2008py,Degrande:2011ua}, including a muon collider-like detector simulation with \verb"Delphes"~\cite{deFavereau:2013fsa}.
To train the hyperplane separating the events from the signal and the background, the events from the SM contribution and aQGCs contribution are generated, the interference terms between the SM and aQGCs are ignored in the training phase.
In the training phase, the events are generated with the same basic cut as the default.
The cuts w.r.t. the infrared divergence are,
\begin{equation}
\begin{split}
&p_{T,\gamma} > 10\;{\rm GeV},\;\; |\eta _{\gamma}| < 2.5, \;\; \Delta R_{\gamma\gamma} > 0.4,
\end{split}
\label{eq.standardcuts}
\end{equation}
where $p_{T,\gamma}$ is the transverse momentum of the photon and $\eta _{\gamma}$ is the pseudo-rapidity of the photon, $\Delta R_{\gamma\gamma}=\sqrt{\Delta \phi ^2 + \Delta \eta ^2}$ where $\Delta \phi$ and $\Delta \eta$ are difference between the azimuth angles and pseudo-rapidity of two photons, respectively.
The events for signals are generated by one operator at a time, with the coefficients chosen as the upper bounds listed in Table~\ref{table:1}.

\begin{figure}[htpb]
\centering
\begin{center}
\includegraphics[width=0.48\hsize]{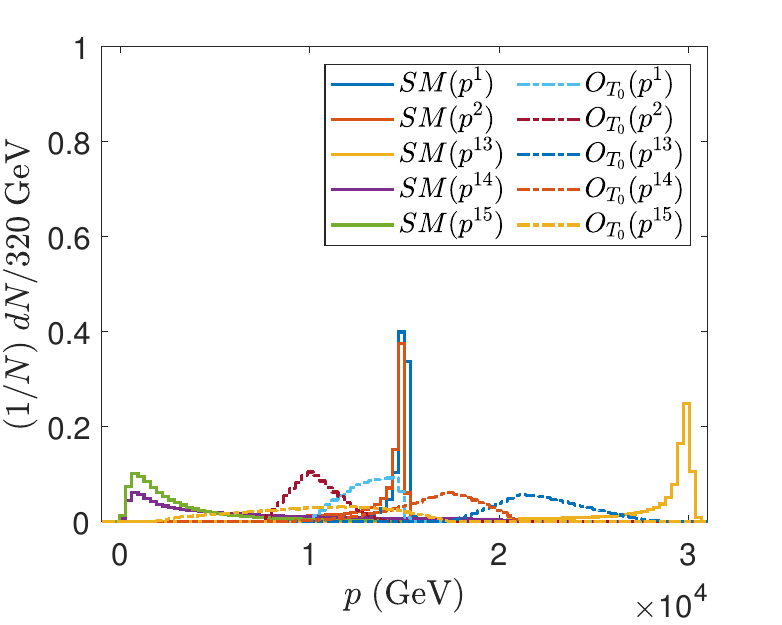}
\includegraphics[width=0.48\hsize]{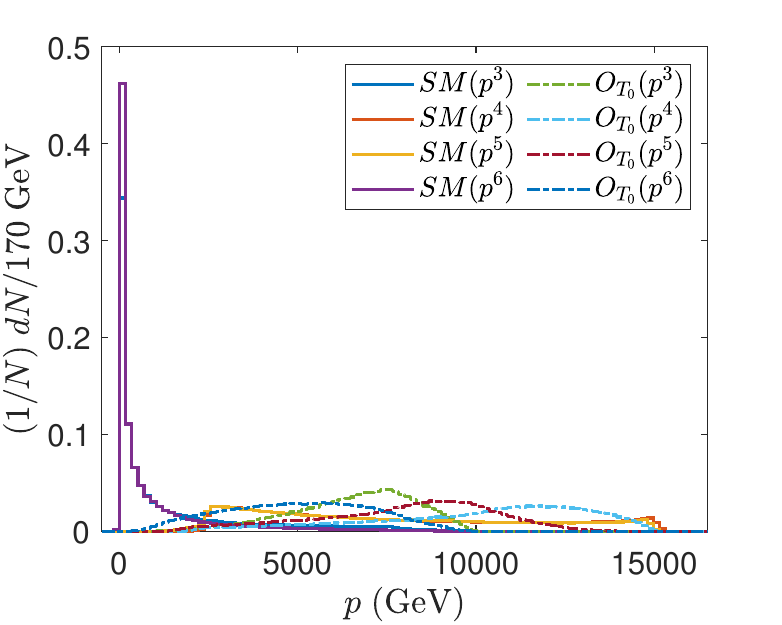}
\includegraphics[width=0.48\hsize]{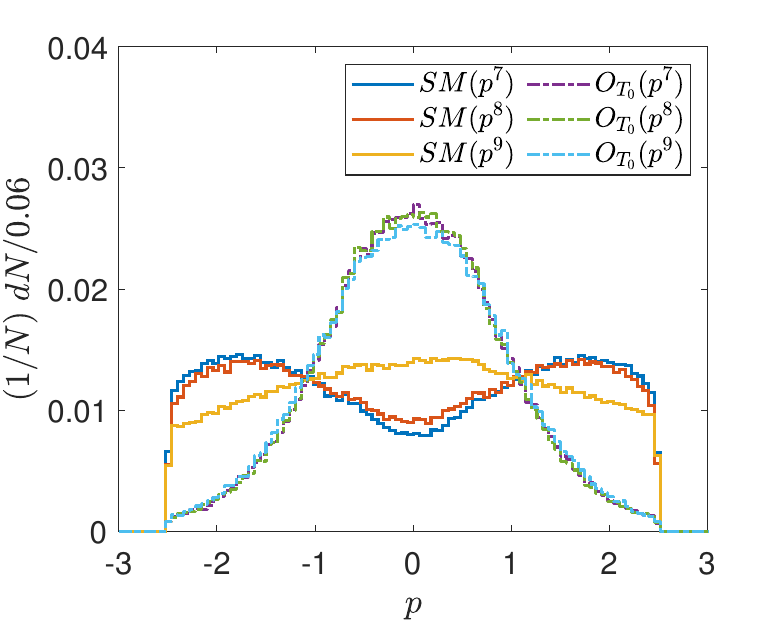}
\includegraphics[width=0.48\hsize]{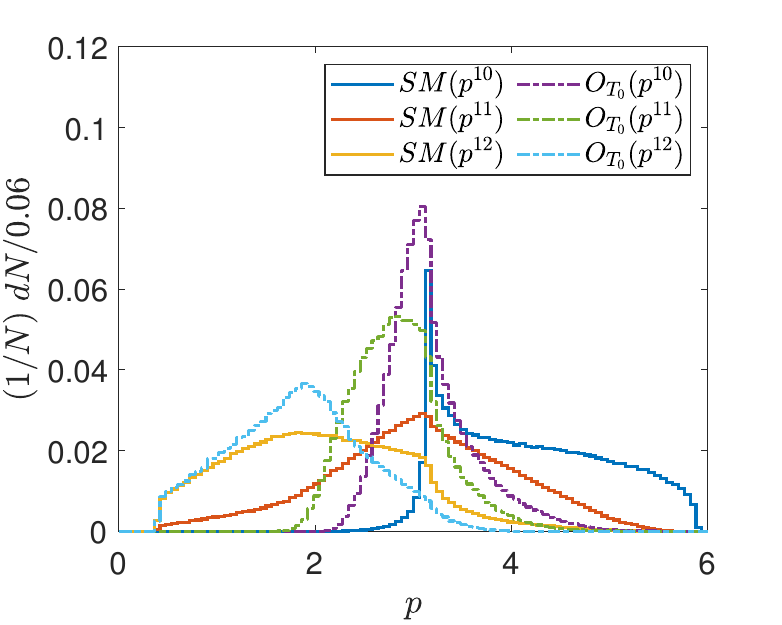}
\caption{\label{fig:feature}
Normalized distributions of $p^j$ for the events from the SM and $O_{T_0}$ contributions at $\sqrt{s}=30\;\rm{TeV}$.}
\end{center}
\end{figure}
In the following, each event is required to consist of at least three photons. 
In this paper, the three hardest photons are chosen and the following observables are calculated, the energies $E$ of the photons, the transverse momenta $p_{T}$ of the photons, the pseudo-rapidities $\eta$ of the photons, $\Delta R$ and the invariant mass $m$ between every two photons.
Denoting the index of the photons according to their energy order, for example $E_{1,2,3}$ represent the energies of the hardest, second hardest and third hardest photons respectively.
These observables form a $15$-dimensional vector denoted $\vec{p}$, where the components $p^j$ are listed in Table~\ref{table:12}.
Fig.~\ref{fig:feature} shows the normalized distributions of $15$ features~($p^j$) for the events from the SM and $O_{T_0}$ contributions at center-of-mass~(c.m.) energy $\sqrt{s}=30\;\rm{TeV}$.
Since SVM finds the separating hyperplane automatically, i.e., SVM does not need to analyze the processes as well as know the physical meanings behind the features, in the following the features and the processes are not analyzed, and we only rely on SVM to get the separating hyperplane.
In the following, $p_{i}^{j}$ is used to denote the $j$-th feature of the $i$-th vector, and we focus on classification using these vectors. 
In the training phase, the interference between SM and NP will be considered in the next section.

\subsection{\label{sec3.3}Using SVM to search for aQGCs}

The soft margin SVM used in this paper is implemented in the \verb"scikit-learn"~\cite{scikit-learn} package.
The data preparation is performed by \verb"MLAnalysis"~\cite{MLAnalysis}.
Before training, the data sets are standardized using the parameters of the the SM training data sets, which will be described in detail in following sections.

The penalty parameter is an important hyper-parameter in the SVM to balance the model complexity and training error. 
In general, the larger the penalty parameter, the harsher the penalty for misclassification and the higher the model complexity, which can lead to overfitting.
The whole analysis is ran for different $c$ and the one that gave the tightest constraint on the expected coefficient is chosen.
The results are presented in the next section.
After comparing the penalty parameter among $c=5$, $10$, $20$, $50$ and $100$, it is found that $c=10$ works best, which is the one chosen in this paper.
In the SVM, one uses $y=\sum _j \left(w^j x^j+b\right)$ to predict the classification of a vector $\vec{x}$, i.e., the classification of $\vec{x}$ is decided by the sign of $y$ and $y>0$ is a criterion such that a point can be identified as more likely from the NP.
In training, to obtain the hyperplane, an equal number of the SM and NP events are used, making it best to separate the two classes.
However, at the colliders the SM contribution is dominant, while the NP contribution should be small.
To keep enough NP signals, instead of $y>0$, we use $y>y_{th}$ to select the events, with the threshold $y_{th}$ maximizes the signal significance.

To summarizes, the steps to use the SVM to optimize the event selection strategy is listed as follows,
\begin{enumerate}
\item Using MC to generate two training data sets, which contain the SM events and the NP events.
\item Select a set of observables which will be used by the event selection strategy, map each event to a point in the multidimensional space where the axes are the observables in the set. 
\item Assign labels to the points from both training data sets, such that $-1$ is for the SM and $+1$ is for the NP, after z-score standardization, use the SVM to calculate the hyperplane, i.e. to obtain the parameters $w^j$ and $b$.
\item With a test data set which could be from the experiment or from MC with interference between the SM and NP included, calculate $y_i=\sum _j \left(w^j x_i^j+b\right)$ where $x^j$ are components of the $i$-th point in the test data set but standardized using parameters from the SM training data set.
\item Use $y_i>y_{th}$ as a cut to select the events, where $y_{th}$ is chosen to maximize the signal significance.
\end{enumerate}

Note that, $y_i>y_{th}$ corresponds to an explicit event selection strategy $\sum_j w^j x^j +b > y_{th}$, where $w^j$, $b$ and $y_{th}$ are constant numbers in the point view of the test data set, and $x^j$ are observables standardized using parameters from the SM training data set, therefore $x^j$ are observables standardized using constant numbers in the point view of the test data set.
As will be shown in the next section, in practice usage, we change $y_{th}$ gradually until we find the value that maximize the signal significance.

\section{\label{sec4}Numerical results of the SVM}

\begin{table}[htbp]
\centering
\begin{tabular}{c|c|c|c|c|c} 
\hline
  $\sqrt{s}({\rm TeV})$ & $ w^{1}$& $ w^{2} $ & $ w^{3} $ & $ w^{4}$& $w^{5}$\\
  \hline
 $3$ & $-1.712$& $-2.379$&$-1.003$&$0.0256$&$0.192$\\
  \hline
 $10$ & $-1.653$&$-2.441$&$-0.772$&$0.0213$&$0.157$\\
  \hline
 $14$ &$-1.630$&$-2.318$&$-1.544$&$0.0709$&$0.192$\\
  \hline
 $30$ & $-1.476$& $-2.220$&$-1.529$&$0.0404$&$0.192$\\
  \hline
  $\sqrt{s}({\rm TeV})$ & $ w^{6}$& $ w^{7} $ & $ w^{8} $ & $ w^{9}$& $w^{10}$ \\ 
  \hline
  $3$ & $0.269$&$ 0.00184$&$0.00855$&$0.0215$&$-0.194$\\
  \hline
 $10$ &$ 0.237$&$0.0285$&$0.0228$&$0.0142$& $-0.224$\\
  \hline
 $14$&$0.279$&$-0.00625$&$-0.0138$&$-0.0120$& $-0.173$\\
  \hline
 $30$ & $0.254$&$ 0.0304$&$0.0223$&$0.00710$&$-0.154$\\
  \hline
  $\sqrt{s}({\rm TeV})$ & $ w^{11}$& $ w^{12} $ & $ w^{13} $ & $ w^{14}$& $w^{15}$ \\ 
  \hline
 $3$ &$-0.253$&$-0.326$&$3.812$&$ 2.412$&$1.069$\\
  \hline
 $10$ & $-0.272$& $-0.319$& $4.065$&$2.592$&$1.055$\\
  \hline
 $14$ & $-0.257$& $-0.252$& $3.570$&$3.104$&$1.098$\\
  \hline
 $30$ &$-0.266$&$-0.334$&$3.407$&$ 3.106$&$1.243$\\
  \hline
\end{tabular}
\caption{$w^j$ obtained at $\sqrt{s}= 3\;{\rm TeV}$, $ 10\;{\rm TeV}$, $ 14\;{\rm TeV}$, and $ 30\;{\rm TeV}$, respectively.}
\label{table:w}
\end{table}

\begin{table}[htbp]

\centering
\begin{tabular}{c|c|c|c|c} 
\hline
  $\sqrt{s}$& $ 3\;{\rm TeV}$ & $ 10\;{\rm TeV}$ & $ 14\;{\rm TeV}$  & $ 30\;{\rm TeV}$  \\ 
 \hline
$b$ &$ -0.555$&$-0.748$&$-0.864$&$ -0.933$\\
 \hline
\end{tabular}
\caption{$b$ obtained at $\sqrt{s}= 3\;{\rm TeV}$, $10\;{\rm TeV}$,  $14\;{\rm TeV}$, and $30\;{\rm TeV}$, respectively.}
\label{table:b}
\end{table}

\begin{figure}[htpb]
\begin{center}
\includegraphics[width=0.8\hsize]{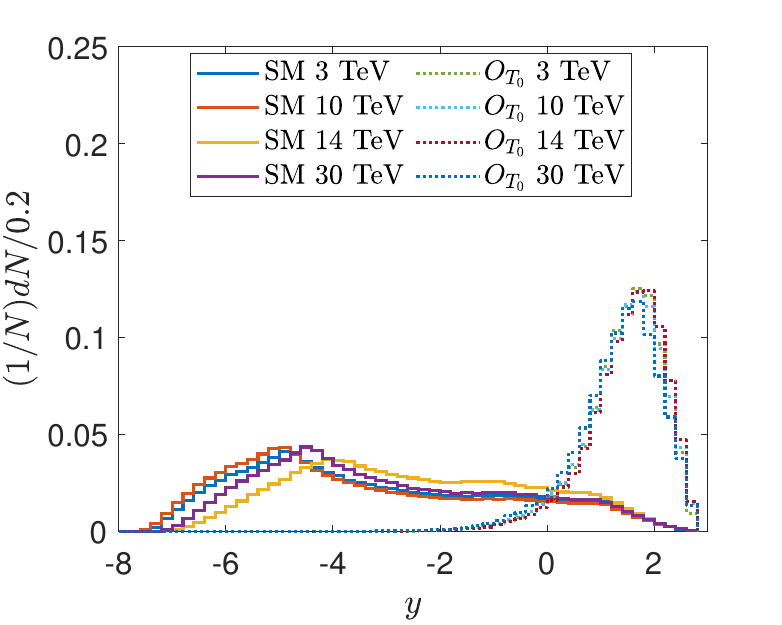}
\caption{\label{fig:distributionofy}
The normalized distributions of $y$ for the SM background and NP signals at different c.m. energies.}
\end{center}
\end{figure}

Before training, a z-score standardization~\cite{Donoho_2004} is used, which calculates the mean values and the standard deviations, and then use, 
\begin{equation}
\begin{split}
x_{i}^{j} =\frac{p_{i}^{j} - \bar{p}^{j}}{d^j},
\end{split}
\label{eq.zscore}
\end{equation} 
to replace $p_i^j$, where $\bar{p}^{j}$ and $d^j$ are the mean values and the standard deviations of the $j$-th features over the SM training data sets at different c.m. energies.
$\bar p^{j}$ and $d^{j}$ are listed in \ref{sec:ap1}.

In the training phase, we use $30000$ NP events and $30000$ SM events.
After training, the $w^j$ and $b$ can be obtained, which are listed in Tables~\ref{table:w} and \ref{table:b}.
In the following, to verify that there is no overfitting, we use validation data sets which use different events from the training data sets.
The validation data sets consist of $600000$ events from the SM, and $100000$ events from the $O_{T_{0}}$ operators.
The normalized distributions of $y^{SM}$ and $y^{NP}$ are shown in Fig.~\ref{fig:distributionofy}, where $y^{SM}$ and $y^{NP}$ are $y$ calculated with the SM and the NP events from the validation data sets, respectively.
It can be seen in Fig.~\ref{fig:distributionofy} that the distributions of $y^{SM}$ and $y^{NP}$ peak at different positions, indicating that $y$ can be used for discriminating the NP signals from the SM background.

\subsection{\label{sec4.1}Expected constraints on the coefficients}

\begin{table}[htbp]
\centering
\begin{tabular}{c|c|c|c} 
\hline
  $\sqrt{s}$&Unit of coefficient&$f_{T_0}/\Lambda ^4$&$f_{T_2}/\Lambda ^4$\\
  \hline
  $3\;{\rm TeV}$&$({\rm TeV^{-4}})$&$[-0.3, 0.3]$&$[-0.5, 0.5]$ \\ 
  \hline
  $10\;{\rm TeV}$&$(10^{-3}\;{\rm TeV^{-4}})$&$[-1.5, 1.5]$&$[-2.0, 2.0]$\\ 
  \hline
  $14\;{\rm TeV}$& $(10^{-3}\;{\rm TeV^{-4}})$&$[-0.5, 0.5]$&$[-0.8, 0.8]$\\ 
  \hline
  $30\;{\rm TeV}$&$ (10^{-4}\;{\rm TeV^{-4}})$&$[-0.5, 0.5]$&$[-0.8, 0.8]$ \\ 
 \hline
   $\sqrt{s}$&Unit of coefficient&$f_{T_5}/\Lambda ^4$&$f_{T_7}/\Lambda ^4$\\
  \hline
  $3\;{\rm TeV}$&$({\rm TeV^{-4}})$&$[-0.06, 0.06]$&$[-0.1, 0.1]$ \\ 
  \hline
  $10\;{\rm TeV}$&$(10^{-3}\;{\rm TeV^{-4}})$&$[-0.3, 0.3]$&$[-0.5, 0.5]$\\ 
  \hline
  $14\;{\rm TeV}$& $(10^{-3}\;{\rm TeV^{-4}})$&$[-0.1, 0.1]$&$[-0.15, 0.15]$\\ 
  \hline
  $30\;{\rm TeV}$&$ (10^{-4}\;{\rm TeV^{-4}})$&$[-0.08, 0.08]$&$[-0.15, 0.15]$ \\ 
  \hline
    $\sqrt{s}$&Unit of coefficient&$f_{T_8}/\Lambda ^4$&$f_{T_9}/\Lambda ^4$\\
  \hline
  $3\;{\rm TeV}$&$({\rm TeV^{-4}})$&$[-0.01, 0.01]$&$[-0.016, 0.016]$ \\ 
  \hline
  $10\;{\rm TeV}$&$(10^{-3}\;{\rm TeV^{-4}})$&$[-0.05, 0.05]$&$[-0.08, 0.08]$\\ 
  \hline
  $14\;{\rm TeV}$& $(10^{-3}\;{\rm TeV^{-4}})$&$[-0.015, 0.015]$&$[-0.02, 0.02]$\\ 
  \hline
  $30\;{\rm TeV}$&$ (10^{-4}\;{\rm TeV^{-4}})$&$[-0.015, 0.015]$&$[-0.02, 0.02]$ \\ 
  \hline
\end{tabular}
\caption{The range of coefficients used in scanning.}
\label{table:coefficients}
\end{table}

When the NP signals cannot be found, the task is to set constraints on the operator coefficients.
To estimate the expected constraints at the muon colliders, in this section, we generate events with the SM contribution, the NP contribution and the interference between the SM and NP all included.
The events are generated assuming one operator at a time, with the range of operator coefficients listed in Table~\ref{table:coefficients}.
For a sufficiently small coefficient, the interference is always more important.
Ignoring the logarithms caused by the phase space integration, and assuming that all scales are negligible compared to $\sqrt{s}$, the dimensional analysis of the cross-section leads to $\sigma _{int}\sim sf_{T_i}/\Lambda ^4$ and $\sigma _{NP}\sim s^3f^2_{T_i}/\Lambda ^8$, where $\sigma _{\rm int}$ and $\sigma _{\rm NP}$ correspond to the contribution of the interference and the contribution of only NP, respectively.
When $\sigma _{\rm int}\approx \sigma _{\rm NP}$, $f_{T_i}/\Lambda ^4 \sim 1/s^2$.
The range of coefficients considered in this paper is larger than the above estimate.
Taking $O_{T_0}$ as an example, $\sigma _{int}=f_{T_0}e^4ss_W^2\left(284\log(2)-215\right)/\left(27648\pi^3\Lambda ^4\right)$, $\sigma _{\rm NP}=f_{T_0}^2e^2s^3s_W^4/\left(34560\pi^3\Lambda ^4\right)$, for the range of coefficients considered in this paper, at $\sqrt{s}=3\; {\rm TeV}$, $\sigma _{\rm int}$ is of the same order of magnitude as $\sigma _{\rm NP}$, while at $\sqrt{s}=30\;{\rm TeV}$ the $\sigma _{\rm NP}$ is two orders of magnitude larger.

It has been shown that $p_{T,\gamma}$ cut can effectively suppress the SM backgrounds~\cite{Yang:2020rjt}, in order to relieve the pressure on computing resources, in the standard cut we use $p_{T,\gamma} > 0.1 E_{\rm beam}$, where $E_{\rm beam}$ is the beam energy, and the other cuts w.r.t. the infrared divergence are as same as in Eq.~(\ref{eq.standardcuts}).

\begin{figure}[htpb]
\begin{center}
\includegraphics[width=0.48\hsize]{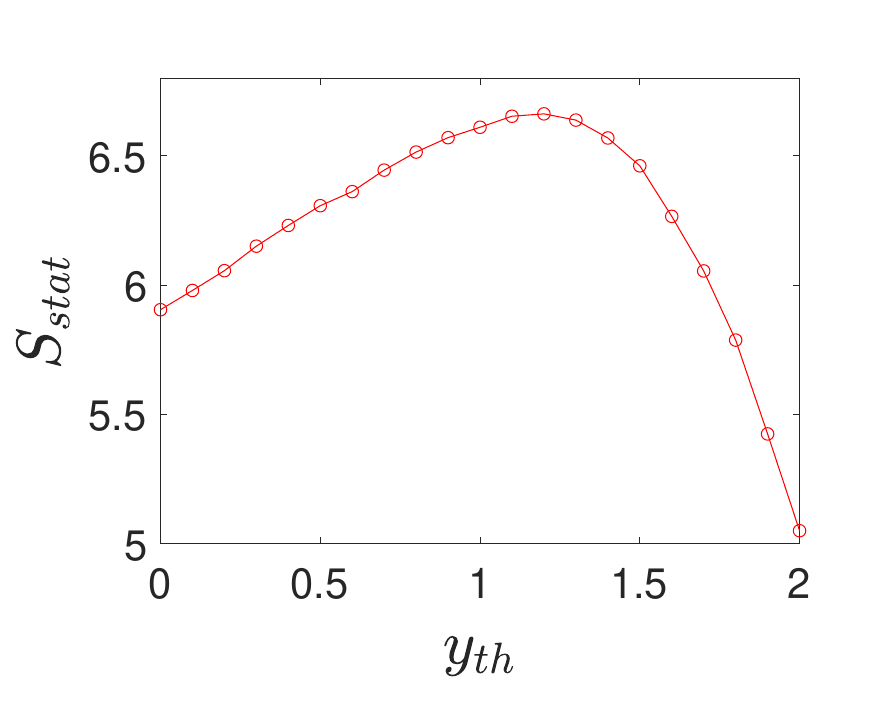}
\includegraphics[width=0.48\hsize]{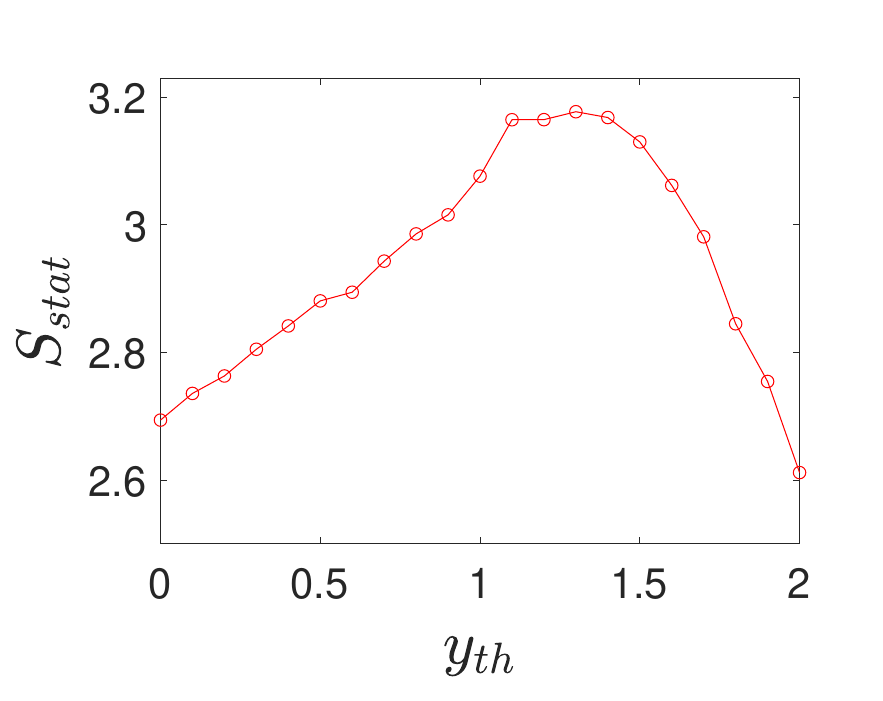}
\includegraphics[width=0.48\hsize]{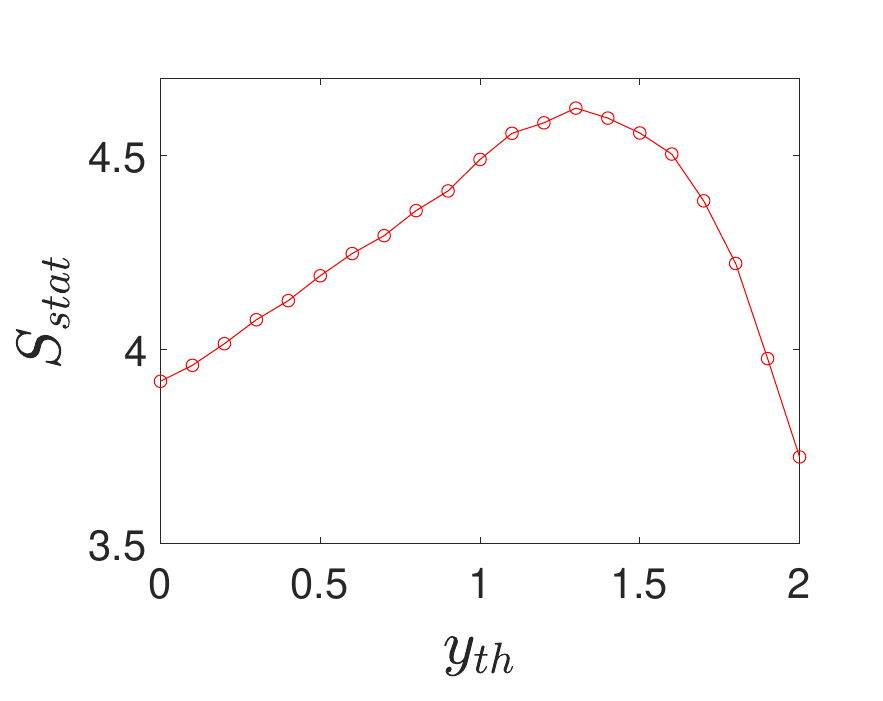}
\includegraphics[width=0.48\hsize]{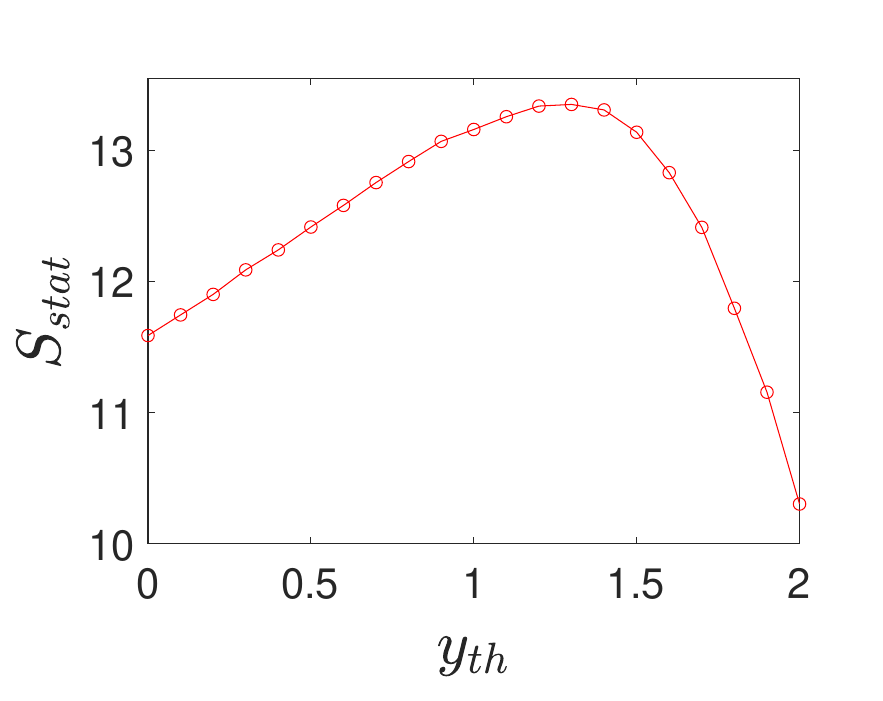}
\caption{\label{fig:ss}
The signal significances as functions of $y_{th}$ for $O_{T_0}$ when the coefficients are picked as the upper bounds listed in Table~\ref{table:coefficients},  and the luminosities are picked as in the ``conservative'' case.
The top-left panel corresponds to $\sqrt{s} = 3\;{\rm TeV}$, the top-right panel corresponds to $\sqrt{s} =10\;{\rm TeV}$, the bottom-left panel corresponds to $\sqrt{s} =14\;{\rm TeV}$, and the bottom-right panel corresponds to $\sqrt{s} =30\;{\rm TeV}$.}
\end{center}
\end{figure}

\begin{table}[htbp]
\centering
\begin{tabular}{c|c|c|c|c} 
\hline
  $\sqrt{s}$& $ 3\;{\rm TeV}$ & $ 10\;{\rm TeV}$ & $ 14\;{\rm TeV}$  & $ 30\;{\rm TeV}$  \\ 
 \hline
 $y_{th}$ & $1.2$&$1.3$&$1.3$&$1.3$\\
 \hline
\end{tabular}
\caption{The $y_{th}$ picked in this paper at $\sqrt{s}= 3\;{\rm TeV}$, $10\;{\rm TeV}$,  $14\;{\rm TeV}$, and $30\;{\rm TeV}$, respectively.}
\label{table:yth}
\end{table}

In order to optimize the $y_{th}$, the cross-sections with $f_{T_i}$ equal to the upper bounds listed in Table~\ref{table:coefficients} are considered.
The statistical signal significance after cuts is calculated which is defined as~\cite{Cowan:2010js,10.1093/ptep/ptaa104}
\begin{equation}
\begin{split}
&\mathcal{S}_{stat}=\sqrt{2 \left[(N_{\rm bg}+N_{s}) \ln (1+N_{s}/N_{\rm bg})-N_{s}\right]},
\end{split}
\label{eq.ss}
\end{equation} 
where $N_s=(\sigma-\sigma_{\rm SM})L$ and $N_{\rm bg}=\sigma_{\rm SM}L$, where $\sigma$ is the cross-section after cuts, $\sigma _{\rm SM}$ is the contribution of the SM after cuts, and $L$ representing the luminosity.
In the ``conservative'' case~\cite{AlAli:2021let}, $L=1\;{\rm ab}^{-1}$, $10\;{\rm ab}^{-1}$, $10\;{\rm ab}^{-1}$ and $10\;{\rm ab}^{-1}$ at $\sqrt{s}=3\;{\rm TeV}$, $10\;{\rm TeV}$, $14\;{\rm TeV}$ and $30\;{\rm TeV}$, respectively. 
$\mathcal{S}_{stat}$ as functions of $y_{th}$ are calculated, and shown in Fig.~\ref{fig:ss}.
The $y_{th}$ chosen to maximizes the $S_{stat}$ are listed in Table~\ref{table:yth}.
After the $y_{th}$ are fixed, the optimized event selection strategy can be obtained as, $b+\sum _i w^i\left(p^i-\bar{p}^i\right)/d^i > y_{th}$, where $w^i$, $b$ and $y_{th}$ are constant numbers listed in Tables~\ref{table:w}, \ref{table:b} and \ref{table:yth}, respectively, $\bar{p}^i$ and $d^i$ are constant numbers listed in \ref{sec:ap1}, $p^i$ is an observable of an event to be determined, which can be from the experiment.

\begin{figure*}[htpb]
\begin{center}
\includegraphics[width=0.3\textwidth]{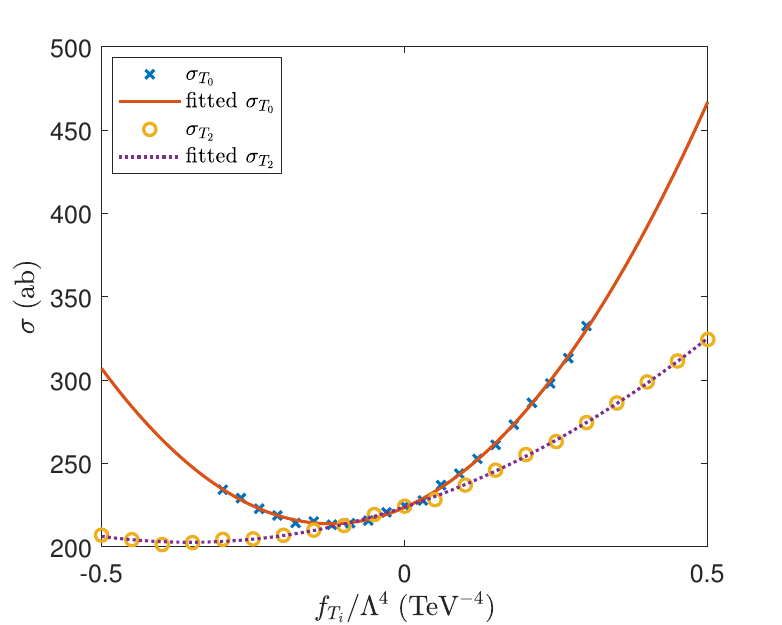}
\includegraphics[width=0.3\textwidth]{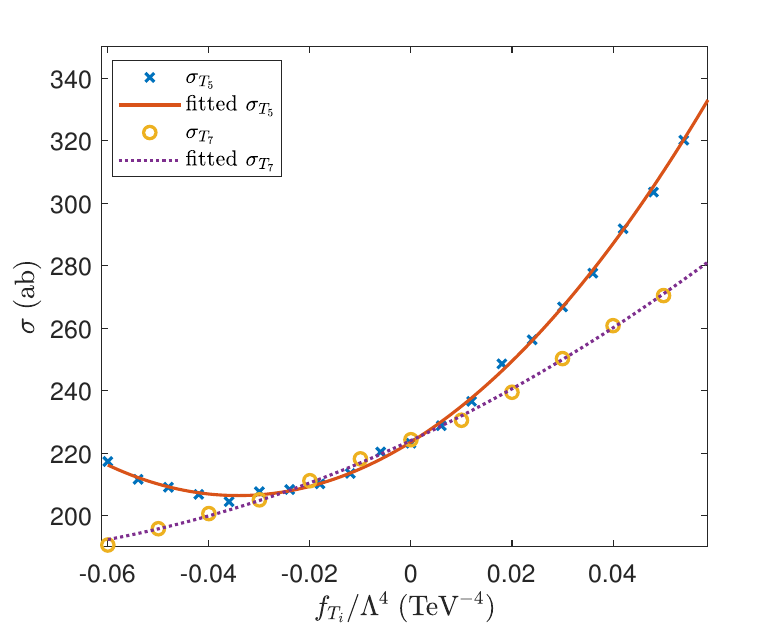}
\includegraphics[width=0.3\textwidth]{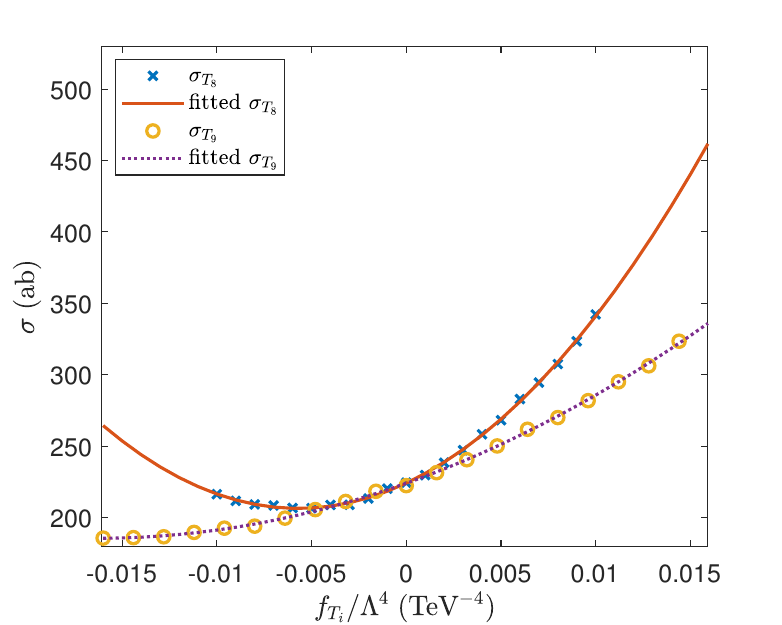}
\includegraphics[width=0.3\textwidth]{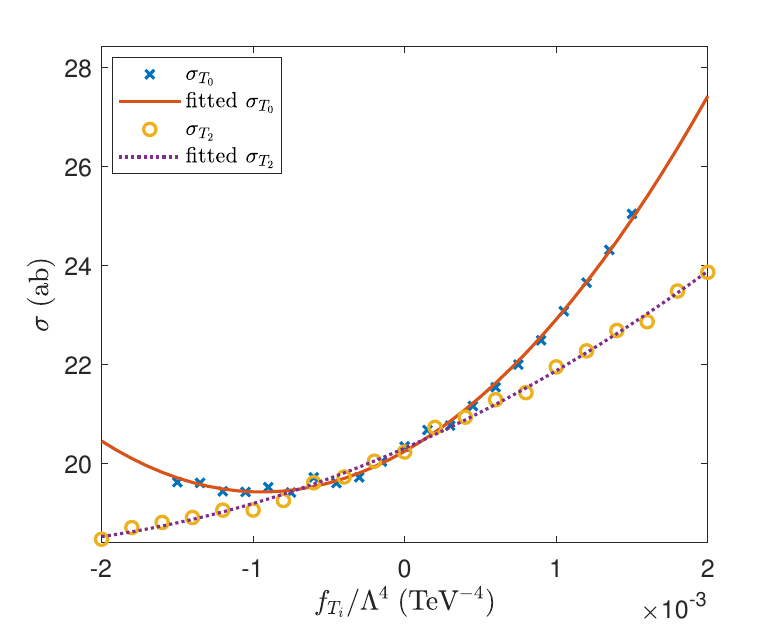}
\includegraphics[width=0.3\textwidth]{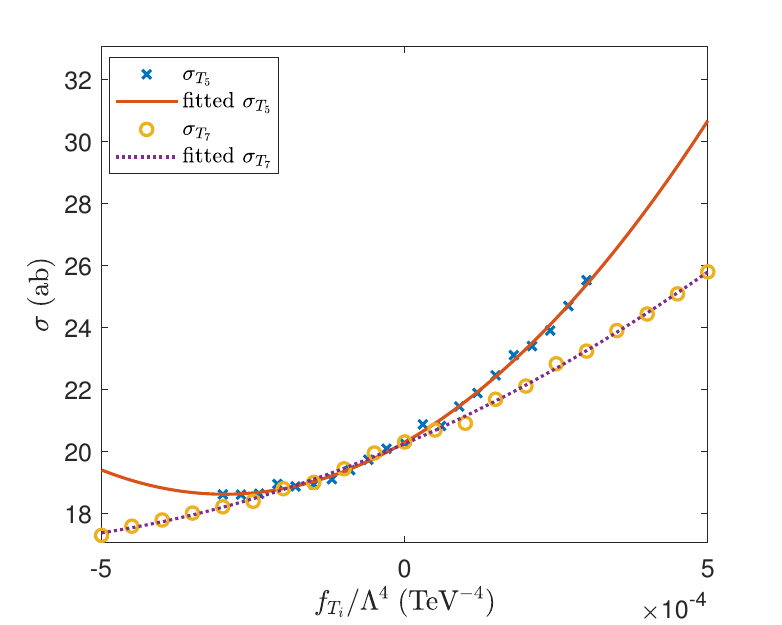}
\includegraphics[width=0.3\textwidth]{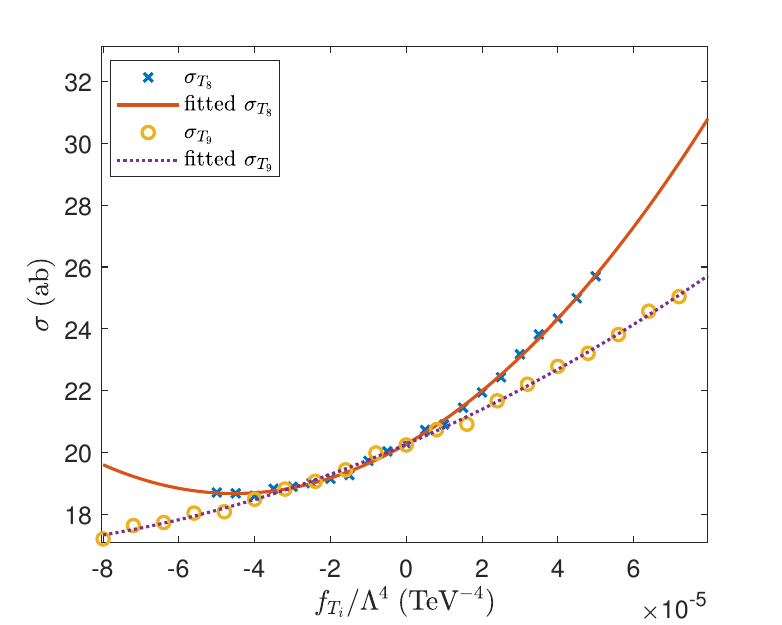}
\includegraphics[width=0.3\textwidth]{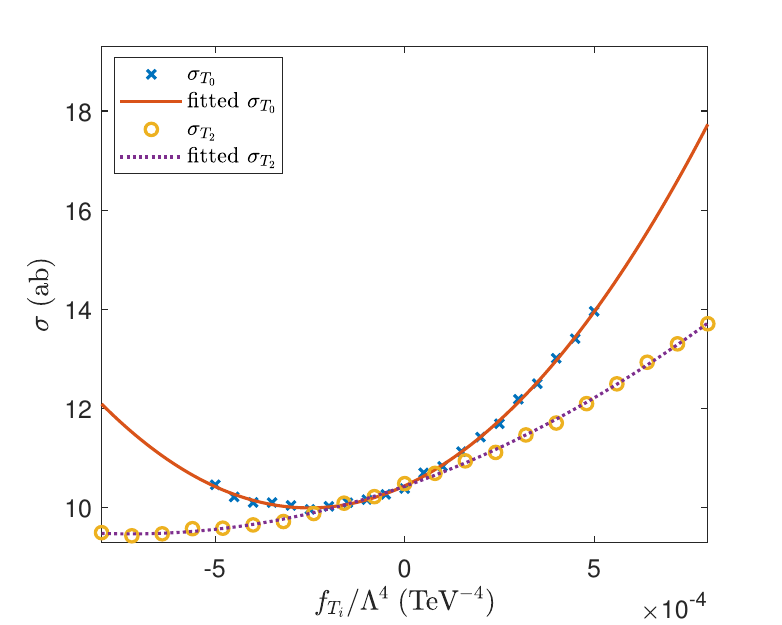}
\includegraphics[width=0.3\textwidth]{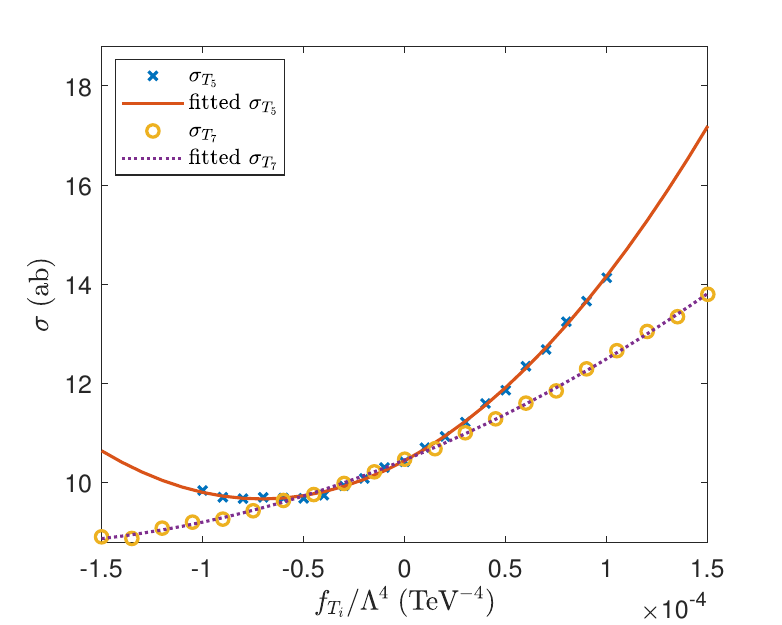}
\includegraphics[width=0.3\textwidth]{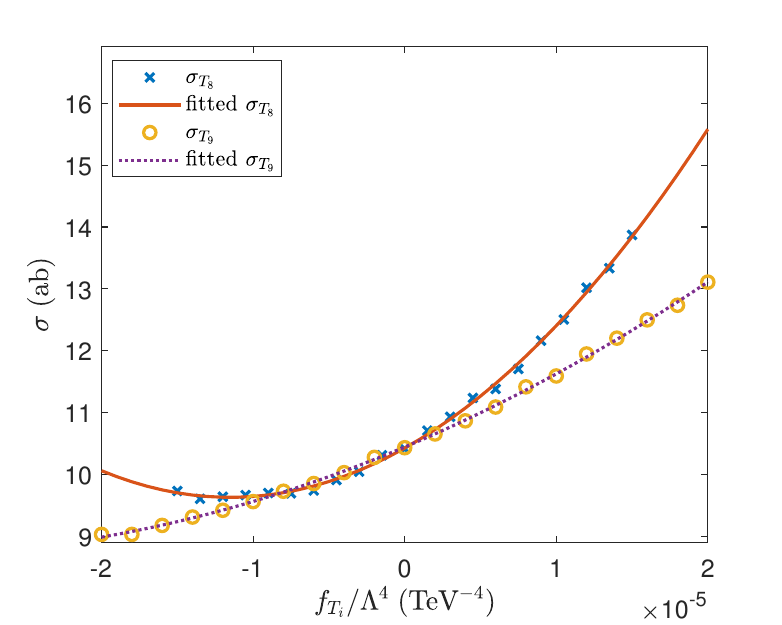}
\includegraphics[width=0.3\textwidth]{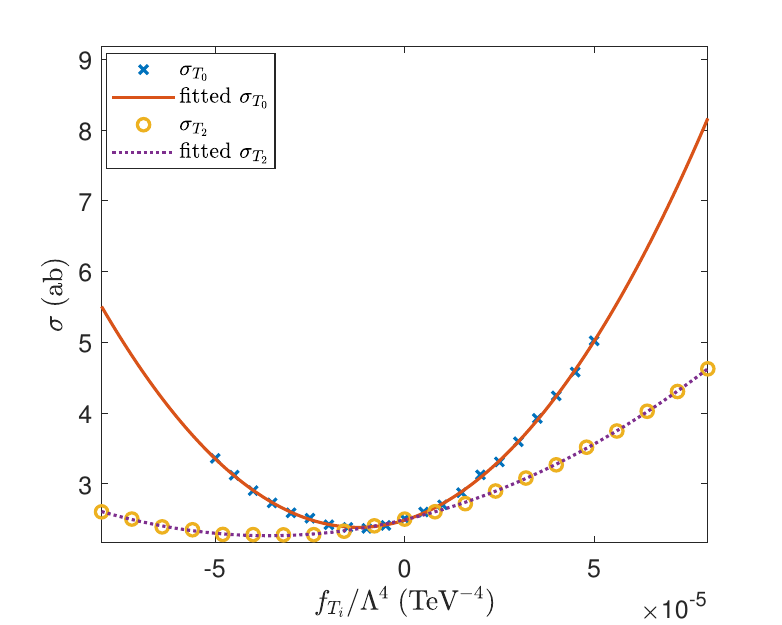}
\includegraphics[width=0.3\textwidth]{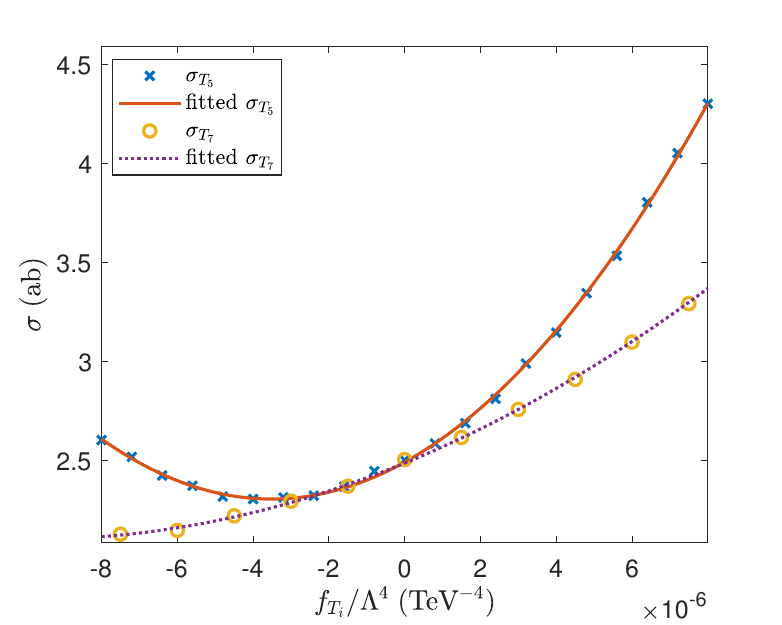}
\includegraphics[width=0.3\textwidth]{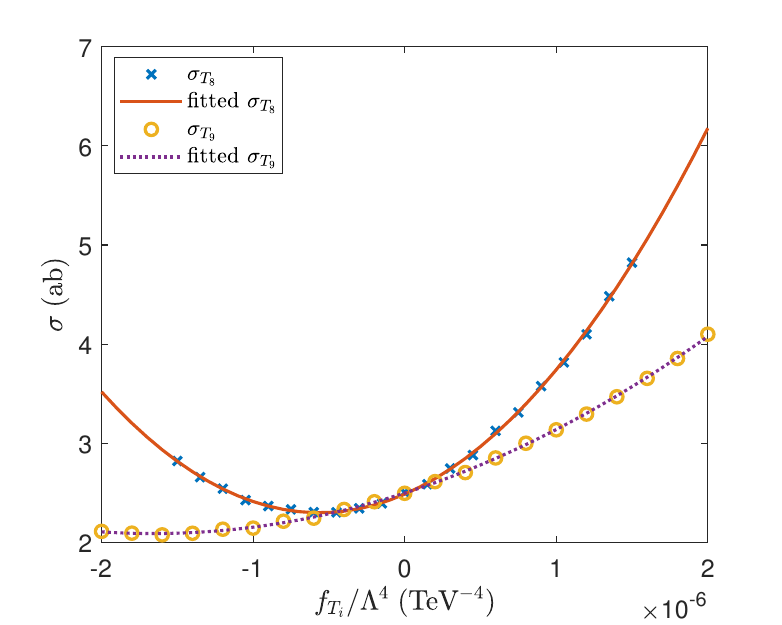}
\caption{\label{fig:cross-section}
The cross-sections $\sigma$ after cuts compared with the fitted $\sigma$ according to bilinear functions of $f_{T_{i}}/\Lambda^4$ at different collision energies, $\sqrt{s} = 3\;{\rm TeV}$ (row 1), $10\;{\rm TeV} (row 2)$, $14\;{\rm TeV}$  (row 3), and $30\;{\rm TeV}$  (row 4).}
\end{center}
\end{figure*}

After applying the event selection strategy, the total cross-section of the tri-photon process with aQGCs can be written as,
\begin{equation}
\begin{split}
&\sigma=\sigma_{\rm SM}+ \hat{\sigma}_{\rm int} f_{T_{i}}/\Lambda^{4}+ \hat{\sigma}_{\rm NP} \left(f_{T_{i}}/\Lambda^{4}\right)^2,
\end{split}
\label{eq.fit}
\end{equation}
where $\sigma _{\rm int} = \hat{\sigma}_{\rm int} f_{T_{i}}/\Lambda^{4}$ and $\sigma _{\rm NP} = \hat{\sigma}_{\rm NP} \left(f_{T_{i}}/\Lambda^{4}\right)^2$.
The results after cuts are shown in Fig.~\ref{fig:cross-section}.
It can be seen that the bilinear function fits well.
From Fig.~\ref{fig:cross-section} it can also be seen that at small c.m. energies the interference between SM and NP is important.
The $y_{th}$ and $c$ have already partially accounted for the interference term.
In feature space, the SVM actually projects the events onto the normal of the hyperplane and then distinguishes the signal events from the background.
In other words, it can be seen as a dimension reduction from the feature space to a one-dimensional space~(which is $y$) and then searching for the signal events in this one-dimensional space.
The effect of the interference is then an increasing or decreasing density of the distribution in overlapping regions of the SM and NP, i.e. about $y\in [-2, 2]$ in Fig.~\ref{fig:distributionofy}.
While $c$ reflects the importance of the overlapping region on the normal of the hyperplane, a non-zero $y_{th}$ to maximize signal significance reflects whether it is better to include or exclude the events in the overlapping region.
However, if the characteristics of $\sigma _{\rm int}$ could be taken into account directly during training, one might expect a better result to be archived, especially for scenarios with smaller $\sqrt{s}$.
From the form of the event selection strategy obtained, there must be cuts specific to the changes in the final state particle distribution, i.e. observables w.r.t. the change in distribution density must be included to make up the feature space, which is worth exploring in the future.

\begin{table*}[htbp]
  \centering
  \begin{tabular}{c|c|c|c|c|c} 
  \hline
   &  &$3\;{\rm TeV} $&$10\;{\rm TeV}$ &$14\;{\rm TeV}$ &$30\;{\rm TeV} $\\
    &$\mathcal{S}_{stat}$& $1\;{\rm ab}^{-1}$&$10\;{\rm ab}^{-1}$&$10\;{\rm ab}^{-1}$ &$10\;{\rm ab}^{-1}$ \\
    &  &$(10^{-2}\;{\rm TeV^{-4}})$&$(10^{-4}\;{\rm TeV^{-4}})$&$(10^{-4}\;{\rm TeV^{-4}})$&$(10^{-5}\;{\rm TeV^{-4}})$ \\
  \hline
    & $2$ &$[-37.08, 12.62]$&$[-29.56, 10.68]$& $[-8.56, 3.52]$&$[-5.36, 2.92]$\\
  $f_{T_{0}}(f_{T_{1}})/\Lambda^4$& $3 $&$[-41.53, 17.07]$&$[-33.27, 14.39]$& $[-9.75, 4.71]$&$[-6.28, 3.84]$ \\
    &$ 5 $& $[-49.05, 24.59]$&$[-39.52, 20.64]$&$[-11.75, 6.70]$&$[-7.85, 5.40]$\\
  \hline
   & $2$ &$[-91.06, 20.06]$&$[-76.88, 16.96]$&$[-20.21, 5.70]$&$[-12.16, 4.98]$ \\
  $f_{T_{2}}/\Lambda^4$&$ 3$ &$[-98.98, 27.97]$&$[-83.58, 23.67]$&$[-22.35, 7.85]$&$[-13.90, 6.73]$ \\
    & $5$ &$[-112.77, 41.76]$&$[-95.27, 35.35]$&$[-26.06, 11.55]$&$[-16.90, 9.72]$ \\
  \hline
   & $2 $&$[-9.12, 2.29]$&$[-7.89, 1.95]$&$[-2.07, 0.658]$& $[-1.26, 0.559]$\\
  $f_{T_{5}}(f_{T_{6}})/\Lambda^4$&$ 3 $&$[-10.00, 3.16]$&$[-8.63,2.69]$&$[-2.31, 0.898]$&$[-1.45, 0.750]$ \\
    & $5$ &$[-11.51, 4.67]$&$[-9.93, 3.99]$&$[-2.72, 1.31]$&$[-1.78, 1.08]$ \\
  \hline
  & $2$ &$[-5.27, 3.47]$&$[-4.70, 2.91]$&$[-5.22,1.03]$& $[-2.89, 0.923]$\\
  $f_{T_{7}}/\Lambda^4$& $3$ &$[-25.10, 4.95]$&$[-19.89, 4.15]$&$[-5.64, 1.45]$&$[-3.24, 1.27]$ \\
    &$ 5$ &$[-27.76, 7.61]$&$[-22.10, 6.35]$&$[-6.38, 2.19]$& $[-3.85, 1.88]$ \\
  \hline
   &$2$ &$[-1.51, 0.371]$&$[-1.22, 0.309]$&$[-0.336,  0.105]$&$[-0.202, 0.0891]$ \\
  $f_{T_{8}}/\Lambda^4$& $3$ &$[-1.65, 0.514]$&$[-1.34, 0.427]$&$[-0.374, 0.143]$&$[-0.232, 0.120]$ \\
    &$ 5$ &$[-1.90, 0.761]$&$[-1.54, 0.631]$&$[-0.440,  0.209]$& $[-0.284, 0.172]$\\
  \hline
   &$2$ &$[-0.832, 0.554]$&$[-0.732, 0.470]$&$[-0.820, 0.165]$&$[-0.481, 0.149]$ \\
  $f_{T_{9}}/\Lambda^4$& $3 $&$[-4.05, 0.791]$&$[-3.31, 0.670]$&$[-0.913, 0.232]$& $[-0.538, 0.206]$\\
    & $5$ &$[-4.48, 1.22]$&$[-3.67, 1.03]$&$[-1.03, 0.353]$&$[-0.637, 0.305]$\\
  \hline
  \end{tabular}
  \caption{In the "conservative" case, the projected sensitivities on the coefficients of the $O_{T_i}$ operators at the muon colliders for various c.m. energies and integrated luminosities.}
  \label{table:8}
  \end{table*}
    
  \begin{table}[htbp]
  \centering
  \begin{tabular}{c|c|c|c} 
  \hline
    &  &$14\;{\rm TeV}$ &$30\;{\rm TeV}$ \\
    &$\mathcal{S}_{stat}$& $20\;{\rm ab}^{-1}$&$90\;{\rm ab}^{-1}$ \\
    &  &$(10^{-5}\;{\rm TeV^{-4}})$&$(10^{-6}\;{\rm TeV^{-4}})$ \\
  \hline
    & $2$ &$[-77.65, 27.20]$&$[-37.71, 13.26]$ \\
  $f_{T_{0}}(f_{T_{1}})/\Lambda^4$& $3$ &$[-87.18, 36.74]$&$[-42.35, 17.90]$ \\
    & $5 $& $[-103.26, 52.81]$& $[-50.16, 25.71]$ \\
  \hline
   & $2$ &$[-187.98, 42.94]$&$[-92.70, 20.91]$ \\
  $f_{T_{2}}/\Lambda^4$&$ 3$ &$[-204.81, 59.76]$&$[-100.91, 29.12]$ \\
    &$ 5 $&$[-234.07, 89.02]$& $[-115.18, 43.39]$ \\
  \hline
   & $2 $&$[-19.13, 5.00]$&$[-9.43, 2.40]$ \\
  $f_{T_{5}}(f_{T_{6}})/\Lambda^4$&$ 3 $&$[-21.03, 6.89]$&$[-10.34, 3.31]$ \\
    & $5 $& $[-24.29, 10.16]$&$[-11.92, 4.89]$ \\
  \hline
  & $2$ &$[-11.99, 7.59]$&$[-5.95, 3.66]$ \\
  $f_{T_{7}}/\Lambda^4$& $3$ &$[-52.70, 10.81]$&$[-24.86, 5.20]$ \\
    & $5 $&$[-58.47, 16.58]$& $[-27.61, 7.96]$ \\
  \hline
   &$2 $&$[-3.10, 0.796]$&$[-1.51, 0.381]$ \\
  $f_{T_{8}}/\Lambda^4$&$ 3 $&$[-3.41, 1.10]$&$[-1.65, 0.527]$ \\
    &$5$ &$[-3.93, 1.62]$& $[-1.90, 0.778]$ \\
  \hline
   &$2 $&$[-1.90, 1.22]$&$[-2.40, 0.588]$ \\
  $f_{T_{9}}/\Lambda^4$&$ 3 $&$[-8.54, 1.73]$&$[-4.16, 0.836]$ \\
    & $5 $&$[-9.47, 2.66]$&$[-4.60, 1.28]$ \\
  \hline
  \end{tabular}
  \caption{Same as Table~\ref{table:8} but for the ``optimistic'' case.}
  \label{table:9}
  \end{table}

To estimate the expected constraints on the coefficients, both the luminosities in the ``conservative'' and ``optimistic'' cases are considered~\cite{AlAli:2021let}.
With the help of $\mathcal{S}_{stat}$ and the fitted cross-sections after cuts, the expected constraints on the operator coefficients at $2\sigma$, $3\sigma$, and $5\sigma$ levels are calculated and presented in Tables~\ref{table:8} and \ref{table:9}.

\begin{table}[htbp]
\centering
\begin{tabular}{c|c|c|c} 
\hline
   & & &$ 30\;{\rm TeV}$ \\
   operator&$c$&$S_{stat}$&$10\;{\rm ab}^{-1}$  \\
   & & &$ (10^{-5}\;{\rm TeV^{-4}})$ \\
  \hline
   $f_{T_0}$&$5$&2& $[-5.36069,2.91584]$ \\
  \hline
   $f_{T_0}$&$10$&2& $[-5.35886,2.91536]$ \\  
  \hline
   $f_{T_0}$&$20$&2& $[-5.36135,2.91582]$ \\  
  \hline
   $f_{T_0}$&$50$&2& $[-5.36142,2.91567]$ \\  
  \hline
   $f_{T_0}$&$100$&2&$[-5.36058,2.91594]$ \\  
  \hline
\end{tabular}
\caption{Expected coefficient constraints on $f_{T_0}$ at $\sqrt{s}=30\;{\rm TeV}$ and $\mathcal{S}_{stat}=2$ in ``conservative case'' when $c=5$, $10$, $20$, $50$ and $100$, respectively.}
\label{table:c}
\end{table}
Applying the analysis with different $c$, we find that the effect of penalty parameter $c$ is small, which is shown in Table~\ref{table:c}, indicating that $c=10$ works best.

\subsection{\label{sec4.2}Simplified event selection strategy}

When the amount of data to be processed is very large, the operational efficiency of the event selection strategy is also a very important consideration. 
As can be seen from Table~\ref{table:w}, some of the $w^j$ are much smaller than the others. 
From the point of view of the feature space, this suggests that the best separation hyperplane is almost parallel to the axes corresponding to these observables. 
As a result, these observables play a relatively minor role in the event selection strategy. 
When computing power is an important factor, one can ignore these observables with very small $w$, i.e., only uses those observables with large $w$ to form a simplified event selection strategy. 
In this paper, a simplified event selection strategy is considered that uses only $6$ observables, $\sum_{j=1,2,3,13,14,15} w^j ({p_i^j - \bar{p}^j})/{d^j} +b > y_{th}^s$ (they are $E_1,E_2,E_{3}, m_{12},m_{13},m_{23}$) and is denoted as `simplified' event selection strategy.

\begin{table}[htbp]
\centering
\begin{tabular}{c|c|c|c|c} 
\hline
  $\sqrt{s}$& $ 3\;{\rm TeV}$ & $ 10\;{\rm TeV}$ & $ 14\;{\rm TeV}$  & $ 30\;{\rm TeV}$  \\ 
 \hline
 $y_{th}$ & $0.9$&$0.8$&$0.8$&$0.9$\\
 \hline
\end{tabular}
\caption{Same as Table~\ref{table:yth} but for the `simplified' event selection strategy.}
\label{table:yths}
\end{table}

\begin{table*}[htbp]
\centering
\begin{tabular}{c|c|c|c|c|c} 
\hline
  &  &$3\;{\rm TeV}$ &$10\;{\rm TeV}$ &$14\;{\rm TeV}$ &$30\;{\rm TeV} $\\
  &$S_{stat}$& $1\;{\rm ab}^{-1}$&$10\;{\rm ab}^{-1}$&$10\;{\rm ab}^{-1}$ &$10\;{\rm ab}^{-1}$ \\
  &  &$(10^{-2}\;{\rm TeV^{-4}})$&$(10^{-4}\;{\rm TeV^{-4}})$&$(10^{-4}\;{\rm TeV^{-4}})$&$(10^{-5}\;{\rm TeV^{-4}})$ \\
\hline
  & $2$ &$[-39.23, 15.04]$&$[-32.99, 12.77]$& $[-9.19,4.12]$&$[-5.82, 3.34]$\\
$f_{T_{0}}(f_{T_{1}})/\Lambda^4$& $3 $&$[-44.33, 20.14]$&$[-37.29, 17.07]$& $[-10.53, 5.46]$&$[-6.84, 4.36]$ \\
  & $5 $& $[-52.82, 28.63]$&$[-44.45, 24.23]$&$[-12.74, 7.67]$&$[-8.54, 6.06]$\\
\hline
 &$ 2 $&$[-95.41, 24.39]$&$[-78.95, 20.27]$&$[-21.43, 6.84]$&$[-13.14, 5.83]$ \\
$f_{T_{2}}/\Lambda^4$& $3$ &$[-104.66, 33.63]$&$[-83.54, 27.85]$&$[-23.88, 9.30]$&$[-15.10, 7.78]$ \\
  &$ 5 $&$[-120.51, 49.48]$&$[-96.48, 40.80]$&$[-28.05, 13.46]$&$[-18.40, 11.08]$ \\
\hline
 &$ 2$ &$[-9.60, 2.76]$&$[-7.58, 2.28]$&$[-2.15, 0.768]$& $[-1.37, 0.640]$\\
$f_{T_{5}}(f_{T_{6}})/\Lambda^4$& $3 $&$[-10.61, 3.78]$&$[-8.41, 3.11]$&$[-2.41, 1.04]$&$[-1.58, 0.861]$ \\
  & $5$ &$[-12.34, 5.51]$&$[-9.81, 4.51]$&$[-2.87, 1.48]$&$[-1.93, 1.22]$ \\
\hline
& $2$ &$[-10.08, 4.31]$&$[-7.65, 3.50]$&$[-5.28, 1.22]$& $[-3.10, 1.08]$\\
$f_{T_{7}}/\Lambda^4$& $3$ &$[-26.23, 6.07]$&$[-20.22, 4.92]$&$[-5.76, 1.69]$&$[-3.49, 1.47]$ \\
  & $5$ &$[-29.35, 9.19]$&$[-22.72, 7.41]$&$[-6.58, 2.52]$& $[-4.16, 2.14]$ \\
\hline
 &$2$ &$[-1.60, 0.447]$&$[-1.24, 0.364]$&$[-0.355,  0.125]$&$[-0.217, 0.103]$ \\
$f_{T_{8}}/\Lambda^4$&$ 3$ &$[-1.77, 0.613]$&$[-1.37, 0.497]$&$[-0.399, 0.168]$&$[-0.251, 0.136]$ \\
  & $5$ &$[-2.04, 0.900]$&$[-1.60, 0.723]$&$[-0.472,  0.242]$& $[-0.307, 0.193]$\\
\hline
 &$2$&$[-1.33, 0.689]$&$[-1.17, 0.568]$&$[-0.897, 0.197]$&$[-0.513, 0.174]$ \\
$f_{T_{9}}/\Lambda^4$& $3$ &$[-4.32, 0.973]$&$[-3.52, 0.802]$&$[-0.974, 0.275]$& $[-0.577, 0.237]$\\
  & $5 $&$[-4.82, 1.48]$&$[-3.93, 1.21]$&$[-1.11, 0.416]$&$[-0.686, 0.347]$\\
\hline
\end{tabular}
\caption{Same as Table~\ref{table:8}, but for the `simplified' event selection strategy.}
\label{table:10}
\end{table*}

\begin{table}[htbp]
\centering
\begin{tabular}{c|c|c|c} 
\hline
  &  &$14\;{\rm TeV}$ &$30\;{\rm TeV}$ \\
  &$S_{stat}$& $20\;{\rm ab}^{-1}$&$90\;{\rm ab}^{-1}$ \\
  &  &$(10^{-5}\;{\rm TeV^{-4}})$&$(10^{-6}\;{\rm TeV^{-4}})$ \\
\hline
  & $2$ &$[-82.90, 32.17]$&$[-40.37, 15.59]$ \\
$f_{T_{0}}(f_{T_{1}})/\Lambda^4$& $3$ &$[-93.72, 42.99]$&$[-45.63, 20.84]$ \\
  &$ 5 $& $[-111.71,60.98]$& $[-54.38, 29.59]$ \\
\hline
 & $2$ &$[-197.90, 52.05]$&$[-98.38,25.21]$ \\
$f_{T_{2}}/\Lambda^4$&$ 3 $&$[-217.44, 71.59]$&$[-107.91, 34.74]$ \\
  & $5 $&$[-250.79, 104.95]$& $[-124.22, 51.06]$ \\
\hline
 & $2 $&$[-19.71, 5.89]$&$[-10.03, 2.86]$ \\
$f_{T_{5}}(f_{T_{6}})/\Lambda^4$&$ 3 $&$[-21.85, 8.03]$&$[-11.08, 3.91]$ \\
  &$ 5$ & $[-25.47, 11.65]$&$[-12.87, 5.70]$ \\
\hline
& $2$ &$[-25.15, 9.08]$&$[-24.66, 4.39]$ \\
$f_{T_{7}}/\Lambda^4$& $3$ &$[-53.43, 12.77]$&$[-26.46, 6.18]$ \\
  & $5$ &$[-59.92, 19.26]$& $[-29.62, 9.35]$ \\
\hline
 &$2$ &$[-3.26, 0.954]$&$[-1.60, 0.453]$ \\
$f_{T_{8}}/\Lambda^4$& $3 $&$[-3.61, 1.30]$&$[-1.76, 0.619]$ \\
  & $5 $&$[-4.20, 1.89]$& $[-2.05, 0.903]$ \\
\hline
 &$2$ &$[-3.08, 1.47]$&$[-1.96, 0.707]$ \\
$f_{T_{9}}/\Lambda^4$& $3$ &$[-9.07, 2.07]$&$[-4.39, 0.998]$ \\
  &$ 5$ &$[-10.13, 3.14]$&$[-4.90, 1.51]$ \\
\hline
\end{tabular}
\caption{Same as Table~\ref{table:10}, but for the ``optimistic'' case.}
\label{table:11}
\end{table}

Note that, in the `simplified' case, we do not choose the six variables to train $\vec{w}$ and $b$ again, but just use $\vec{w}$ and $b$ in Tables~\ref{table:w} and \ref{table:b}.
Following same procedures described in the previous subsection to choose $y_{th}$~(shown in Table~\ref{table:yths}) and fit the cross-sections after cuts, the expected constraints are obtained and presented in the Tables~\ref{table:10} and ~\ref{table:11}.
Comparing these expected constraints with those in Tables~\ref{table:8} and \ref{table:9}, we observe that the `simplified' case is also feasible and can achieve slightly worse results than the full case.

\subsection{\label{sec4.3}A numerical test on non-linear kernel SVM}

A linear kernel is used in this paper for the purpose to obtain an explicit event selection strategy.
A simple, explicit event selection strategy can explicitly show the kinematic features the events to be selected. 
However, using a nonlinear kernel, the SVM can also be used as a classifier itself, at the cost of obtaining an event selection strategy that, if written out explicitly, would be too complicated, and no longer a simple expression from which it would be easy to recognize the corresponding kinematic features of the signal events. 
The benefit gained is that, a nonlinear kernel has the potential to achieve better results with linearly indistinguishable data.

The Gaussian kernel SVM, also known as the radial basis function~(RBF) kernel SVM, is a nonlinear kernel widely used~\cite{Cortes:1995hrp,Scholkopf1996IncorporatingII}. 
As a comparison, the SVM with a Gaussian kernel is used, the kernel function is,
\begin{equation}
\begin{split}
& k(x_i, x_j) = \exp\left(-\gamma \|x_i - x_j\|^2\right),
\end{split}
\label{eq.rbf}
\end{equation}
where $\|x_i - x_j\|$ represents the Euclidean distance between the two feature vectors, $\gamma$ is a parameter that determines the influence of a single training example, and usually set as $1/(n_f v)$ where $n_f$ is the number of features and $v$ is the variance of the feature vectors.

\begin{table}[htbp]
\centering
\begin{tabular}{c|c|c|c|c} 
\hline
  $\sqrt{s}$& $ 3\;{\rm TeV}$ & $ 10\;{\rm TeV}$ & $ 14\;{\rm TeV}$  & $ 30\;{\rm TeV}$  \\ 
 \hline
 $y_{th}$ & $1.2$&$1.1$&$1.2$&$1.1$\\
 \hline
\end{tabular}
\caption{Same as Table~\ref{table:yth} but for the RBF kernel SVM.}
\label{table:rbfyth}
\end{table}

\begin{table*}[htbp]
\centering
\begin{tabular}{c|c|c|c|c|c} 
\hline
    &  &$3\;{\rm TeV}$ &$10\;{\rm TeV} $&$14\;{\rm TeV} $&$30\;{\rm TeV} $\\
    &$S_{stat}$& $1\;{\rm ab}^{-1}$&$10\;{\rm ab}^{-1}$&$10\;{\rm ab}^{-1}$ &$10\;{\rm ab}^{-1}$ \\
    &  &$(10^{-2}\;{\rm TeV^{-4}})$&$(10^{-4}\;{\rm TeV^{-4}})$&$(10^{-4}\;{\rm TeV^{-4}})$&$(10^{-5}\;{\rm TeV^{-4}})$ \\
\hline
    &$ 2 $&$[-36.83, 12.39]$&$[-33.03, 12.16]$& $[-9.43, 4.03]$&$[-5.63, 3.19]$\\
  $f_{T_{0}}(f_{T_{1}})/\Lambda^4$& $3$ &$[--41.44, 18.00]$&$[-37.18, 16.31]$& $[-10.75 5.35]$&$[-6.61, 4.16]$ \\
    &$ 5$ & $[-49.17, 25.73]$&$[-44.08, 23.21]$&$[-12.93, 7.54]$&$[-8.25, 5.80]$\\
\hline
   &$ 2 $&$[-91.52, 21.68]$&$[-79.07, 19.41]$&$[-22.27, 6.65]$&$[-12.81, 5.56]$ \\
  $f_{T_{2}}/\Lambda^4$& $3$ &$[-99.90, 30.06]$&$[-86.46, 26.80]$&$[-24.69, 9.08]$&$[-14.69, 7.44]$ \\
    & $5$ &$[-114.37, 44.53]$&$[-99.11, 39.45]$&$[-28.80, 13.19]$&$[-17.89, 10.63]$ \\
\hline
   & $2 $&$[-9.22, 2.45]$&$[-7.81, 2.21]$&$[-2.23, 0.754]$& $[-1.32, 0.619]$\\
  $f_{T_{5}}(f_{T_{6}})/\Lambda^4$& $3$ &$[-10.14, 3.37]$&$[-8.62, 3.02]$&$[-2.49, 1.02]$&$[-1.53, 0.824]$ \\
    & $5 $&$[-11.72, 4.95]$&$[-9.99, 4.39]$&$[-2.94, 1.46]$&$[-1.87, 1.17]$ \\
\hline
  & $2$ &$[-13.07, 3.77]$&$[-7.83, 3.32]$&$[-5.61, 1.18]$& $[-3.02, 1.03]$\\
  $f_{T_{7}}/\Lambda^4$& $3$ &$[-24.90, 5.35]$&$[-20.34 4.68]$&$[-6.08, 1.65]$&$[-3.39, 1.40]$ \\
    &$ 5 $&$[-27.70, 8.15]$&$[-22.73, 7.07]$&$[-6.89, 2.47]$& $[-4.04, 2.05]$ \\
\hline
   &$2 $&$[-1.47, 0.384]$&$[-1.26, 0.349]$&$[-0.376, 0.122]$&$[-0.212,0.0982]$ \\
  $f_{T_{8}}/\Lambda^4$& $3$ &$[-1.61, 0.530]$&$[-1.39, 0.478]$&$[-0.420, 0.166]$&$[-0.244, 0.131]$ \\
    & $5$ &$[-1.86, 0.781]$&$[-1.61, 0.697]$&$[-0.493, 0.240]$& $[-0.299 0.186]$\\
\hline
   &$2$ &$[-0.915, 0.602]$&$[-1.02, 0.536]$&$[-0.912, 0.189]$&$[-0.500, 0.166]$ \\
  $f_{T_{9}}/\Lambda^4$& $3 $&$[-4.48, 0.857]$&$[-3.41, 0.756]$&$[-0.987, 0.265]$& $[-0.561, 0.227]$\\
    & $5$ &$[-4.93, 1.31]$&$[-3.80, 1.15]$&$[-1.12, 0.396]$&$[-0.667, 0.332]$\\
\hline
\end{tabular}
\caption{Same as Table~\ref{table:8} but for RBF kernel SVM. }
\label{table:rbfc}
\end{table*}
  
\begin{table}[htbp]
\centering
\begin{tabular}{c|c|c|c} 
\hline
    &  &$14\;{\rm TeV}$ &$30\;{\rm TeV}$ \\
    &$S_{stat}$& $20\;{\rm ab}^{-1}$&$90\;{\rm ab}^{-1}$ \\
    &  &$(10^{-5}\;{\rm TeV^{-4}})$&$(10^{-6}\;{\rm TeV^{-4}})$ \\
\hline
    &$ 2 $&$[-85.28, 31.37]$&$[-39.23, 14.77]$ \\
  $f_{T_{0}}(f_{T_{1}})/\Lambda^4$& $3$ &$[-95.98, 42.07]$&$[-44.26, 19.79]$ \\
    &$ 5$ & $[-113.79, 59.87]$& $[-52.64, 28.18]$ \\
\hline
   &$ 2$ &$[-206.58,50.43]$&$[-96.39 23.88]$ \\
  $f_{T_{2}}/\Lambda^4$& $3$ &$[-225.79, 69.64]$&$[-105.51, 33.00]$ \\
    & $5$ &$[-258.70, 102.55]$& $[-121.16, 48.65]$ \\
\hline
   & $2$ &$[-20.49, 5.76]$&$[-9.77, 2.71]$ \\
  $f_{T_{5}}(f_{T_{6}})/\Lambda^4$&$ 3$ &$[-22.61, 7.88]$&$[-10.78, 3.72]$ \\
    & $5 $& $[-26.20 11.47]$&$[-12.50, 5.43]$ \\
\hline
  & $2$ &$[-16.28,8.80]$&$[-8.37, 4.15]$ \\
  $f_{T_{7}}/\Lambda^4$&$ 3 $&$[-56.67, 12.44]$&$[-25.78, 5.86]$ \\
    & $5$ &$[-63.08, 18.85]$& $[-28.80, 8.89]$ \\
\hline
   &$2 $&$[-3.47, 0.934]$&$[-1.56, 0.430]$ \\
  $f_{T_{8}}/\Lambda^4$& $3$ &$[-3.82, 1.28]$&$[-1.72, 0.589]$ \\
    &$ 5$ &$[-4.41, 1.87]$& $[-1.99, 0.862]$ \\
\hline
   &$2$ &$[-2.51, 1.41]$&$[-2.13, 0.668]$ \\
  $f_{T_{9}}/\Lambda^4$& $3$ &$[-9.22, 1.99]$&$[-4.29, 0.945]$ \\
    & $5 $&$[-10.25, 3.02]$&$[-4.78, 1.44]$ \\
\hline
\end{tabular}
\caption{Same as Table~\ref{table:rbfc} but of the ``optimistic'' case.}
\label{table:rbfo}
\end{table}

Taking $c=1$, the number results of $y_{th}$ and the expected constraints in the case of RBF kernel are listed in Table~\ref{table:rbfyth}, Table~\ref{table:rbfc} and Table~\ref{table:rbfo}, respectively.
Compared to Table~\ref{table:8} and Table~\ref{table:9}, the expected constraints on operator coefficients obtained using the RBF kernel are more relaxed. 
However, in comparison to Table ~\ref{table:10} and Table ~\ref{table:11}, the constraints are tighter.

\section{\label{sec5}Numerical result of QSVM}

As the designed luminosities of the colliders increase, so does the volume of data. 
Processing and analyzing such data requires significant computational resources, posing a significant challenge to conventional computers.
Quantum computing, however, has the potential to address these complexities inherent in HEP data analysis, given its ability to handle large data sets and high-dimensional feature spaces.
Although quantum computing is in the era of noisy intermediate-scale quantum (NISQ) devices, the precision tests of the SM provide sufficient motivation for the application of quantum computing to HEP.
Recently, QSVM, quantum variational classifiers and variational quantum algorithms have been studied in the field of HEP~\cite{Guan_2021,Wu:2021xsj,Wu:2020cye,Terashi_2021}.

Quantum computing is a computing paradigm based on quantum mechanics with powerful parallel computing capabilities.
The input and output of classical data to a quantum computer is expensive, so the main motivation for using QSVM in this paper is the potential for future applications. 
In the future, it is possible that the data we need to process will come directly from the quantum, as quantum computers will be able to hold or process much more data simultaneously than classical computers. 
On the other hand, in the case of kernel SVM, for example, it is possible for quantum computers to realize kernel functions that are difficult for classical algorithms to implement~\cite{Havlicek:2018nqz,Sherstov:2020qax,Liu:2020lhd}. 
At the LHC, QSVM has been used to analyze $\bar{t}tH$ production~\cite{Wu:2021xsj} with the kernel function introduced in Ref.~\cite{Havlicek:2018nqz}.
In addition, the kernel function used in this paper can be computed using a swap test, which has the potential to be further optimized~\cite{Fanizza_2020,liu2022multistate}.
Similarly, QSVM has also been used in the analysis of the $e^+e^-\to ZH$ process at the CEPC with an optimized kernel algorithm~\cite{Fadol:2022umw}, concluding that state-of-the-art quantum computing technologies could be utilized by HEP.

In this section we follow the spirit of the previous section, i.e. we use QSVM to find an event selection strategy. 
That is, we will use a kernel that can conveniently form an explicit event selection strategy that can be used classically. 
Of course, the QSVM can also be used directly as a classifier.
As a small step towards using quantum computing to search for NP, our goal is to verify the feasibility of the QSVM.

\subsection{\label{sec5.1}Using QSVM to search for aQGCs}

\begin{figure}[htpb]
\begin{center}
\includegraphics[width=0.8\hsize]{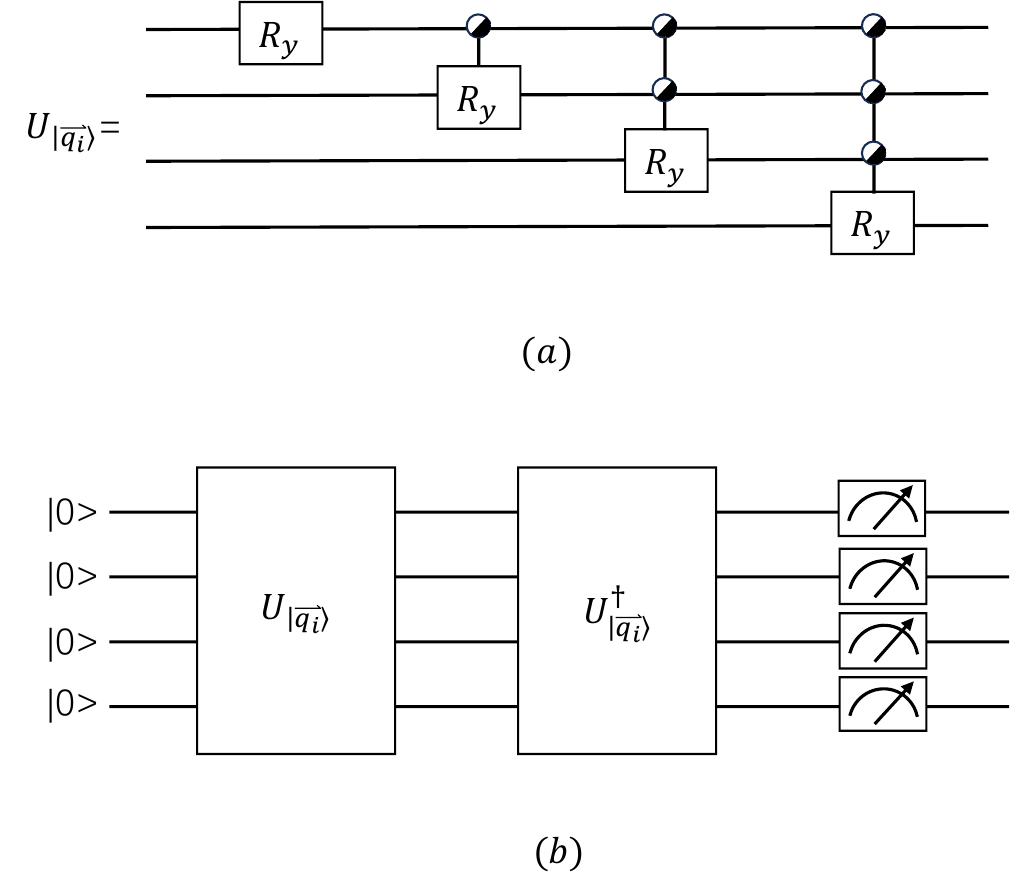}
\caption{\label{fig:circuit}
The circuit for amplitude encode using uniform rotation gates $FR_y$ is depicted in (a). The circuit to calculate $|\langle \vec{q}_i|\vec{q}_j\rangle|$ is depicted in (b) where the probability of the outcome $|0000\rangle$ in the measurement is $|\langle \vec{q}_i|\vec{q}_j\rangle|^2$.}
\end{center}
\end{figure} 
A quantum state can be considered as a normalized complex vector.
However, the classical event selection strategy will be too complicated and not classically practicable if we encode the events as complex vectors directly.
In this section, the events are encoded as normalized real vectors whose components are positive,
\begin{equation}
\begin{split}
&\hat{x}_{i} = \frac{p_{i}-p_{i,{\rm min}}}{p_{i,{\rm max}}-p_{i,{\rm min}}},
\end{split}
\label{eq.x}
\end{equation}
where $p_i$ are data points defined in previous section, $p_{i,{\rm min}}\;(p_{i,{\rm max}})$ is the minimal (maximal) value of the feature $p_{i}$ over the SM training data set, so that most of the values of $\hat{x}_i$ range from $0$ to $1$.
With limited compute resources, we reduce the training data set to $1000$ SM and $1000$ NP events.
The values of $p_{i,{\rm min}}$ and $p_{i,{\rm max}}$ are shown in \ref{sec:ap1}.
A quantum state presenting an event is,
\begin{equation}
\begin{split}
&|\vec{q}\rangle = \frac{1}{\sqrt{\sum _i \hat{x}_i^2 + 1}}\left(|0\rangle + \sum _k\hat{x}_{k}|k\rangle\right),
\end{split}
\label{eq.q}
\end{equation}
where $k$ is the digital representing of a state, i.e. $|0\rangle = |0000\rangle$, $|1\rangle = |0001\rangle$, $|2\rangle = |0010\rangle$, etc.
Using Eq.~(\ref{eq.q}), the length of $\hat{x}$ is also encoded.
A $16$ dimension vector needs four qubits to encode.
In this paper we use amplitude encode and denote $U_{\vec{q}}$ as the circuit such that $U_{\vec{q}}|0\rangle = |\vec{q}\rangle$.
The amplitude encode of Eq.~(\ref{eq.q}) is implemented with the help of uniform rotation gates $FR_y$~\cite{Mottonen:2004vly} as shown in Fig.~\ref{fig:circuit}.~(a), such that $15$ $R_y$ gates and $14$ CNOT gates are used for each vector.

In kernel SVM, the equation to predict the classification is
\begin{equation}
\begin{split}
&y(\vec{q}^{\rm test}_i)=\sum _j\alpha_{j}y_{j}k(\vec{q}^{\rm test}_i,\vec{q}^{\rm sv}_j)+b,\\
\end{split}
\label{eq.kernelsvm}
\end{equation}
where $\vec{q}^{\rm test}_i$ is the vector to be predicted, $\vec{q}^{\rm sv}_j$ are the support vectors found by the SVM, with $y_j$ the labels of $\vec{q}^{\rm sv}_j$ and $\alpha _j$ the weights of $\vec{q}^{\rm sv}_j$, $b$ is a real parameter found by the SVM, and $k$ is the kernel function.

In this paper, we use a linear kernel, $k(\vec{q}^{\rm test}_i,\vec{q}^{\rm sv}_j) =\left| \langle \vec{q}^{\rm sv}|\vec{q}^{\rm test}\rangle\right|=\vec{q}^{\rm sv}_j \cdot \vec{q}^{\rm test}_i$, which can be calculated using a circuit shown in Fig.~\ref{fig:circuit}.~(b).
Note that, two CNOT gates can be cancelled between $U_{\vec{q}_i}$ and $U_{\vec{q}_j}^{\dagger}$, and two $R_y$ gates can be combined.
As a result, $29$ $R_y$ gates and $26$ CNOT gates are used in all.
The probability of the outcome $|0000\rangle$ in the measurements is $|\langle \vec{q}_i|\vec{q}_j\rangle|^2$ in the circuit shown in Fig.~\ref{fig:circuit}.~(b).
In other words, the outcome of the measurement is a list of four numbers, with $2^4$ possible values (each number could be either $0$ or $1$).
When four zeroes are obtained, it indicates that the state of the measured qubits is collapsed to $|0000\rangle$.
The probability that four zeroes are obtained, is $|\langle \vec{q}_i|\vec{q}_j\rangle|^2$.

Note that, $\left| \langle \vec{q}^{\rm sv}|\vec{q}^{\rm test}\rangle\right|=\vec{q}^{\rm sv}_j \cdot \vec{q}^{\rm test}_i$ is ensured by the way we standardize the date set, then Eq.~(\ref{eq.kernelsvm}) is modified as,
\begin{equation}
\begin{split}
&y(\vec{q}^{\rm test}_i)=\left|\vec{w}_q \cdot \vec{q}^{\rm test}_i\right|+b_q,\\
\end{split}
\label{eq.kernelqsvm}
\end{equation}
where $\vec{w}_q = \sum _j\alpha_{j}y_{j} \vec{q}^{\rm sv}_j/\left|\sum _j\alpha_{j}y_{j} \vec{q}^{\rm sv}_j\right|$, $b_q=b/\left|\sum _j\alpha_{j}y_{j} \vec{q}^{\rm sv}_j\right|$.
The normalization factor is added so that the inner product in Eq.~(\ref{eq.kernelqsvm}) can also been calculated using the circuit in Fig.~\ref{fig:circuit}.~(b).
Then, similar as previous section, a tunable parameter $y_{th}$ is introduced to maximize the signal significance, and we use $y(\vec{q}^{\rm test}_i) > y_{th}$ as the event selection strategy.
Note that $y(\vec{q}^{\rm test}_i) > y_{th}$ also corresponds to an explicit event selection strategy.

\subsection{\label{sec5.2}The results of the QSVM event selection strategy}

\begin{table}[htbp]
\centering
\begin{tabular}{c|c|c|c|c} 
\hline
 $\sqrt{s}({\rm TeV})$ & $ w_{q}^{1}$& $ w_{q}^{2} $ & $ w_{q}^{3} $ & $ w_{q}^{4}$\\
 \hline
 $3$ & $-0.0961$& $-0.106$&$0.354$&$0.271$\\
 \hline
 $10$ & $0.0200$&$-0.0899$&$0.338$&$0.223$\\
  \hline
  $14$ &$0.0212$&$-0.127$&$0.342$&$0.334$\\
  \hline
 $30$ &$-0.0322$&$-0.212$&$0.377$&$0.335$\\
  \hline
  $\sqrt{s}({\rm TeV})$ & $ w_{q}^{5}$& $ w_{q}^{6} $ & $ w_{q}^{7} $ & $ w_{q}^{8}$ \\ 
  \hline
  $3$ &$0.213$& $0.498$&$0.122$&$0.0845$\\
 \hline
 $10$ &$0.309$&$0.350$&$0.108$&$0.130$\\
  \hline
 $14$ &$0.173$&$0.441$&$0.0532$& $0.0513$\\
  \hline
 $30$ &$0.223$&$0.395$&$0.0588$&$0.0455$\\
  \hline
  $\sqrt{s}({\rm TeV})$ & $ w_{q}^{9}$& $ w_{q}^{10} $ & $ w_{q}^{11} $ & $ w_{q}^{12}$ \\ 
  \hline
  $3$&$0.158$ &$-0.212$&$-0.184$&$-0.179$\\
 \hline
 $10$ & $0.144$& $-0.140$& $-0.0940$&$-0.174$\\
  \hline
 $14$ &$0.107$&$-0.157$&$-0.177$& $-0.179$\\
  \hline
 $30$ &$0.111$&$-0.158$& $-0.128$&$-0.112$\\
 \hline
 $\sqrt{s}({\rm TeV})$ & $ w_{q}^{13}$& $ w_{q}^{14} $ & $ w_{q}^{15} $ & $ w_{q}^{16}$ \\ 
  \hline
  $3$ &$-0.0741$&$0.401$&$0.359$&$0.187$\\
 \hline
 $10$ & $0.00180$& $0.504$& $0.432$&$0.247$\\
  \hline
 $14$ &$-0.0402$&$0.496$&$0.365$& $0.214$\\
  \hline
 $30$ &$-0.127$&$0.469$& $0.410$&$0.131$\\
 \hline
\end{tabular}
\caption{Same as Table~\ref{table:w} but for QSVM.}
\label{table:qw}
\end{table}

\begin{table}[htbp]
\centering
\begin{tabular}{c|c|c|c|c} 
\hline
  $\sqrt{s}$& $ 3\;{\rm TeV}$ & $ 10\;{\rm TeV}$ & $ 14\;{\rm TeV}$  & $ 30\;{\rm TeV}$  \\ 
 \hline
 $b_{q}$ & $ -0.395$ & $-0.525$ & $-0.439$ & $-0.361$ \\
 \hline
\end{tabular}
\caption{Same as Table~\ref{table:b} but for the QSVM.}
\label{table:qb}
\end{table}

\begin{table}[htbp]
\centering
\begin{tabular}{c|c|c|c|c} 
\hline
  $\sqrt{s}$& $ 3\;{\rm TeV}$ & $ 10\;{\rm TeV}$ & $ 14\;{\rm TeV}$  & $ 30\;{\rm TeV}$  \\ 
 \hline
 $y_{th}$ & $0.10$&$0.10$&$0.20$&$0.17$\\
 \hline
\end{tabular}
\caption{Same as Table~\ref{table:yth} but for the QSVM.}
\label{table:qyth}
\end{table}

\begin{figure}[htpb]
\begin{center}
\includegraphics[width=0.8\hsize]{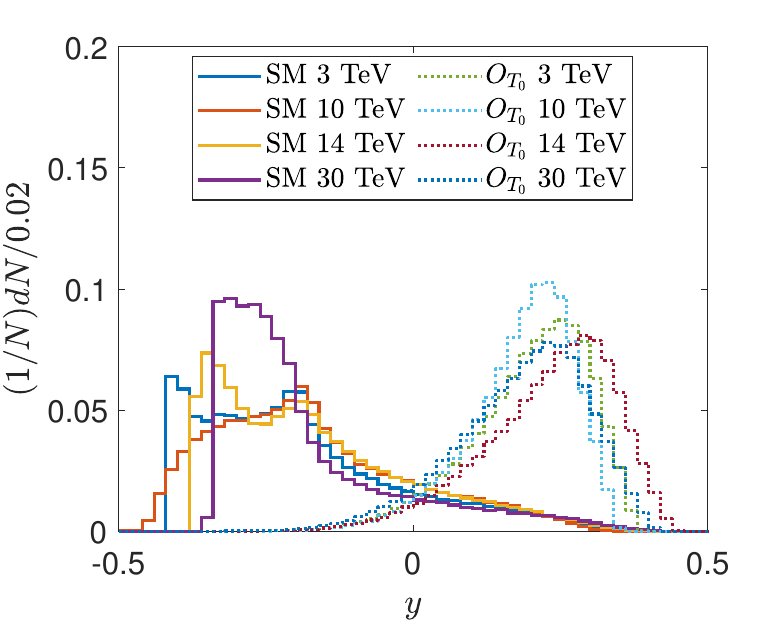}
\caption{\label{fig:qsvmsmnp}
Same as Fig.~\ref{fig:distributionofy} but for the QSVM.}
\end{center}
\end{figure} 
The circuit to calculate the kernel matrix is implemented and simulated using \verb"QuEST"~\cite{Jones_2019}, which is an abbreviation for ``the quantum exact simulation toolkit'', a high performance simulator of quantum circuits, state-vectors and density matrices written in C language supporting GPU acceleration.
The measurement is repeated for $10000$ times for each inner product.
Then the results of the inner products are fed to \verb"scikit-learn" to calculate $\vec{w}_q$ and $b_q$.
After training, the results for the vector $\vec{w}_q$ and real parameter $b_q$ are obtained and shown in Table~\ref{table:qw} and Table~\ref{table:qb}.
Using validation data set consist of $100000$ vectors from the SM, and $100000$ from the NP at each collision energy, the normalized distributions of $y$ are shown in Fig.~\ref{fig:qsvmsmnp}.

\begin{table*}[htbp]
  \centering
  \begin{tabular}{c|c|c|c|c|c} 
  \hline
    &  &$3\;{\rm TeV}$ &$10\;{\rm TeV} $&$14\;{\rm TeV} $&$30\;{\rm TeV} $\\
    &$S_{stat}$& $1\;{\rm ab}^{-1}$&$10\;{\rm ab}^{-1}$&$10\;{\rm ab}^{-1}$ &$10\;{\rm ab}^{-1}$ \\
    &  &$(10^{-2}\;{\rm TeV^{-4}})$&$(10^{-4}\;{\rm TeV^{-4}})$&$(10^{-4}\;{\rm TeV^{-4}})$&$(10^{-5}\;{\rm TeV^{-4}})$ \\
  \hline
    &$ 2 $&$[-36.44, 12.97]$&$[-30.89, 11.23]$& $[-8.45, 3.66]$&$[-5.35, 3.01]$\\
  $f_{T_{0}}(f_{T_{1}})/\Lambda^4$& $3$ &$[-40.94, 17.47]$&$[-34.75, 15.10]$& $[-9.67, 4.88]$&$[-6.29, 3.94]$ \\
    &$ 5$ & $[-48.48, 25.00]$&$[-41.22, 21.57]$&$[-11.74, 6.95]$&$[-7.87, 5.52]$\\
  \hline
   &$ 2 $&$[-93.05, 20.88]$&$[-74.38, 17.53]$&$[-20.12, 6.07]$&$[-12.17, 5.21]$ \\
  $f_{T_{2}}/\Lambda^4$& $3$ &$[-101.21, 29.04]$&$[-81.15, 24.30]$&$[-22.39, 8.34]$&$[-13.96, 7.00]$ \\
    & $5$ &$[-115.36, 43.19]$&$[-92.85, 36.00]$&$[-26.31, 12.26]$&$[-17.02, 10.06]$ \\
  \hline
   & $2 $&$[-9.13, 2.36]$&$[-7.58, 2.02]$&$[-2.11, 0.699]$& $[-1.26, 0.581]$\\
  $f_{T_{5}}(f_{T_{6}})/\Lambda^4$& $3$ &$[-10.03, 3.25]$&$[-8.34,2.78]$&$[-2.36, 0.953]$&$[-1.45, 0.776]$ \\
    & $5 $&$[-11.56, 4.78]$&$[-9.63, 4.08]$&$[-2.80, 1.39]$&$[-1.79, 1.11]$ \\
  \hline
  & $2$ &$[-5.70, 3.60]$&$[-5.32, 3.01]$&$[-4.89,1.08]$& $[-2.86, 0.963]$\\
  $f_{T_{7}}/\Lambda^4$& $3$ &$[-25.31, 5.12]$&$[-19.63, 4.27]$&$[-5.30, 1.51]$&$[-3.22, 1.32]$ \\
    &$ 5 $&$[-28.02, 7.84]$&$[-21.86, 6.50]$&$[-6.07, 2.28]$& $[-3.84, 1.94]$ \\
  \hline
   &$2 $&$[-1.50, 0.381]$&$[-1.20, 0.321]$&$[-0.336,  0.111]$&$[-0.201, 0.0923]$ \\
  $f_{T_{8}}/\Lambda^4$& $3$ &$[-1.65, 0.526]$&$[-1.32, 0.441]$&$[-0.377, 0.151]$&$[-0.232, 0.123]$ \\
    & $5$ &$[-1.89, 0.774]$&$[-1.53, 0.647]$&$[-0.446,  0.221]$& $[-0.285, 0.176]$\\
  \hline
   &$2$ &$[-0.902, 0.575]$&$[-0.841, 0.487]$&$[-0.826, 0.176]$&$[-0.471, 0.155]$ \\
  $f_{T_{9}}/\Lambda^4$& $3 $&$[-4.08, 0.818]$&$[-3.21, 0.690]$&$[-0.898, 0.248]$& $[-0.529, 0.213]$\\
    & $5$ &$[-4.51, 1.25]$&$[-3.57, 1.05]$&$[-1.03, 0.375]$&$[-0.629, 0.313]$\\
  \hline
  \end{tabular}
  \caption{Save as Table~\ref{table:8} but for the QSVM.}
  \label{table:q8}
  \end{table*}
  
  \begin{table}[htbp]
  \centering
  \begin{tabular}{c|c|c|c} 
  \hline
    &  &$14\;{\rm TeV}$ &$30\;{\rm TeV}$ \\
    &$S_{stat}$& $20\;{\rm ab}^{-1}$&$90\;{\rm ab}^{-1}$ \\
    &  &$(10^{-5}\;{\rm TeV^{-4}})$&$(10^{-6}\;{\rm TeV^{-4}})$ \\
  \hline
    &$ 2 $&$[-76.23, 28.32]$&$[-37.30, 13.84]$ \\
  $f_{T_{0}}(f_{T_{1}})/\Lambda^4$& $3$ &$[-85.06, 38.15]$&$[-42.05, 18.59]$ \\
    &$ 5$ & $[-102.65, 54.74]$& $[-50.02, 26.55]$ \\
  \hline
   &$ 2$ &$[-186.34, 45.81]$&$[-91.82, 22.16]$ \\
  $f_{T_{2}}/\Lambda^4$& $3$ &$[-204.13, 63.60]$&$[-100.35, 30.70]$ \\
    & $5$ &$[-235.07, 93.53]$& $[-115.12, 45.47]$ \\
  \hline
   & $2$ &$[-19.37, 5.31]$&$[-9.31, 2.52]$ \\
  $f_{T_{5}}(f_{T_{6}})/\Lambda^4$&$ 3$ &$[-21.38, 7.32]$&$[-10.26, 3.47]$ \\
    & $5 $& $[-24.84, 10.78]$&$[-11.88, 5.09]$ \\
  \hline
  & $2$ &$[-15.11, 7.97]$&$[-11.62, 3.87]$ \\
  $f_{T_{7}}/\Lambda^4$&$ 3 $&$[-49.24, 11.30]$&$[-24.47, 5.48]$ \\
    & $5$ &$[-55.19, 17.26]$& $[-27.33, 8.33]$ \\
  \hline
   &$2 $&$[-3.10, 0.845]$&$[-1.49, 0.400]$ \\
  $f_{T_{8}}/\Lambda^4$& $3$ &$[-3.42, 1.16]$&$[-1.64, 0.551]$ \\
    &$ 5$ &$[-3.97, 1.71]$& $[-1.89, 0.808]$ \\
  \hline
   &$2$ &$[-2.26, 1.30]$&$[-2.08, 0.620]$ \\
  $f_{T_{9}}/\Lambda^4$& $3$ &$[-8.35, 1.85]$&$[-4.04, 0.879]$ \\
    & $5 $&$[-9.34, 2.84]$&$[-4.50, 1.34]$ \\
  \hline
  \end{tabular}
  \caption{Same as Table~\ref{table:q8} but for the ``optimistic'' case.}
  \label{table:q9}
  \end{table}
  
Similar as the previous section, $y_{th}$ are choose to maximize the signal significances.
Then the expected constraints on coefficients are estimated using the same data sets as the previous section but with the event selection strategy $y(\vec{q})>y_{th}$, where $y(\vec{q})$ is defined in in Eq.~(\ref{eq.kernelqsvm}), $y_{th}$ are shown in Table~\ref{table:qyth}.
The expected constraints using QSVM event selection strategy are shown in Tables~\ref{table:q8} and \ref{table:q9}.

When considering the optimization of SVM algorithms for quantum computers, the main consideration is to target the bottlenecks of the whole algorithm that consume the most computational resources, i.e. the computation of the kernel matrix. 
Since we are dealing with classical data in this paper, the swap test is not used, but the use of the swap test will give exactly the same results. 
If we assume that quantum data will be processed in the future, swap test can be used and the optimization is done in two ways.
On the one hand, the swap test does not depend on the length of the vectors whose inner product is to be computed. 
On the other hand, if the number of qubits is large enough, one can use the multi-state swap test~\cite{liu2022multistate}, which has a query number of the order of $\mathcal{O}(n\log(n))$ when the number of events in the training data set is $n$, compared to $\mathcal{O}(n^2)$ times on a classical computer.
The multi-state swap test is appropriate when the kernel is an inner product, for example it can also be used in the case of Refs.~\cite{Wu:2021xsj,Fadol:2022umw}.

Although a linear kernel is used in this paper, the purpose is to obtain an explicit event selection strategy. 
QSVM supports more complex kernels if the purpose is to use QSVM to search for the signal events directly. 
For example, encoding the vector as $|\vec{q}_c\rangle \propto \left[\left(\hat{x}_1+{\rm i}\right)|0\rangle + \sum _k \left(\hat{x}_{2k}+\hat{x}_{2k+1}{\rm i}\right)|k\rangle\right]$, results in fewer qubits~(three qubits) and fewer gates~($28$ $R_{y,z}$ gates and $16$ CNOT gates if we only replace the $U_{\vec{q}_i}$ circuit to encode a complex vector) while archiving a result for $f_{T_0}/\Lambda^4$ at $30\; {\rm TeV}$ and $\mathcal{S}_{stat}=2$, which is $[-5. 35, 3.00]\times 10^{-5}\;{\rm TeV}^{-4}$, which is very close to Table~\ref{table:q8}.
In the following, the results with this kernel is denoted as `complex'.
QSVM supports more complex kernels that require exponential operations on classical computers~\cite{Liu:2020lhd}, which means that for the same amount of data, QSVM has the potential to achieve a further improvement in NP detection than in this paper without exponentially increasing computational resources.

At current stage, the noise of a quantum computer is inevitably involved.
The systematic error of QSVM comes from two main sources, one is to get $w^j$ from the algorithm and exact kernel matrix, and the other is to compute the kernel matrix by a quantum computer.

To evaluate the former, we use the exactly computed kernel matrix and randomly pick $100:100$ vectors, $1000:1000$ vectors, and $10000:10000$ vectors from the SM and NP contributions, respectively, to compute $w^j$ and repeat the computation $100$ times. 
In the case of $\sqrt{s}=30\;{\rm TeV}$ and linear kernel SVM, for $100:100$ vectors, $1000:1000$ vectors, and $10000:10000$ vectors, the relative standard deviations of $w^j$ are $1.74\%$, $1.44\%$, and $0.28\%$, respectively. 
It can be seen that sufficiently large number of vectors will greatly reduce the systematic error of $w^j$.

In the case of QSVM, the error in calculating the kernel matrix comes from two sources, the noise of the quantum computer and the number of measurements.
The effect of noise is studied using density matrix simulation provided by \verb"QuEST".
The noise is simulated by applying a $0.1\%$ depolarizing and a $0.1\%$ dephasing after each gate operation, and a $2\%$ amplitude-damping error to the whole~\cite{Jones_2019,RAUCH1999277}.
If the noise is ignored, the relative standard deviation should not vary with the number of measurements when the number of measurements is large enough. 
We find that in the presence of noise, the relative standard deviation decreases slightly with the number of measurements, i.e., multiple measurements slightly cancel out the error due to noise.
The average fidelity is only about $46\%$, and the standard deviations for kernel matrix elements for $100$, $1000$ and $10000$ measurements are $7.43\%$, $6.25\%$ and $6.10\%$, respectively.

If one simply assumes a linear transfer of relative error, and ignores the errors from the MC simulation, from the finite number of validation events, one can expect the systematic errors of the expected constraints on operator coefficients in the case of the classical and quantum SVM to be about $<0.28\%$~(in the training phase, $30000:30000$ vectors are used therefore the deviation should be smaller than the case of $10000:10000$), and $\pm 1.44\%  \pm 6.10\%$, respectively.

\section{\label{sec6}Comparison of expected coefficient constraints}

\begin{figure*}[htpb]
\begin{center}
\includegraphics[width=0.4\textwidth]{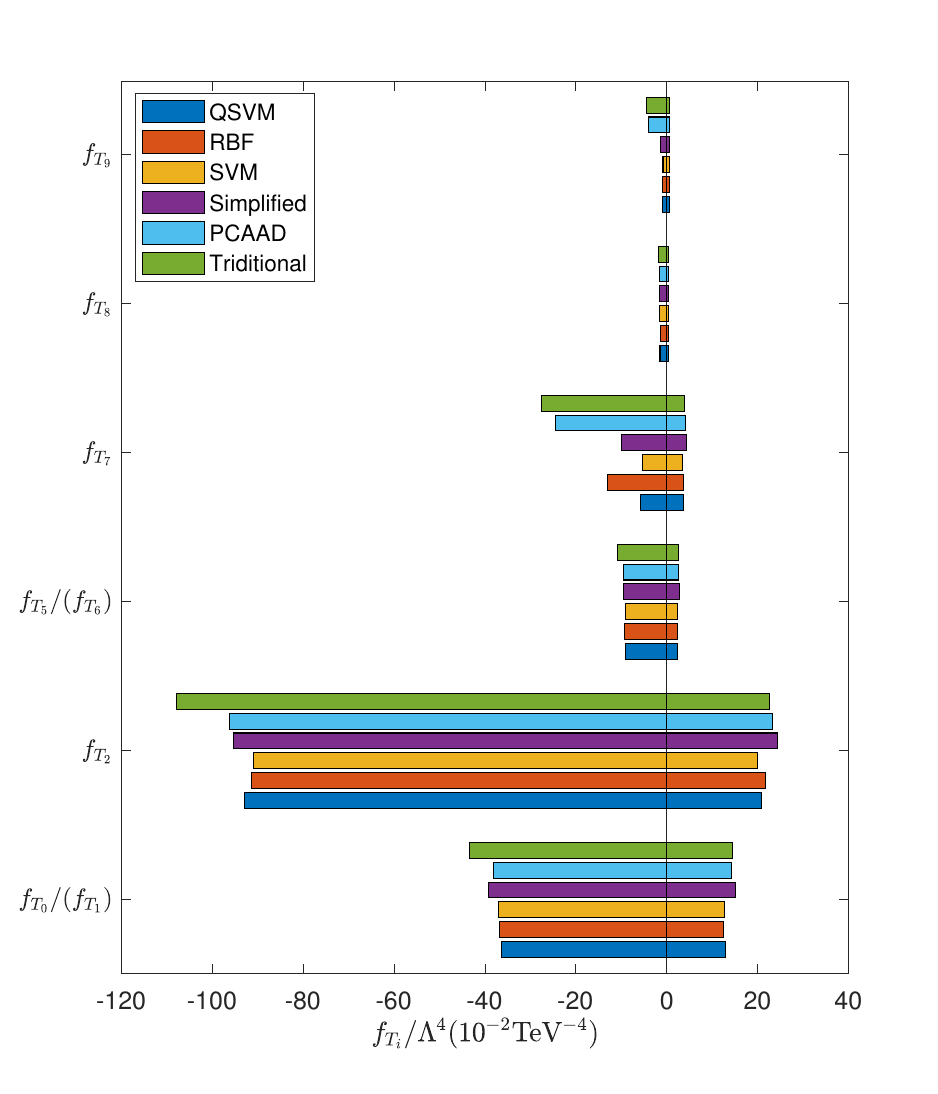}
\includegraphics[width=0.4\textwidth]{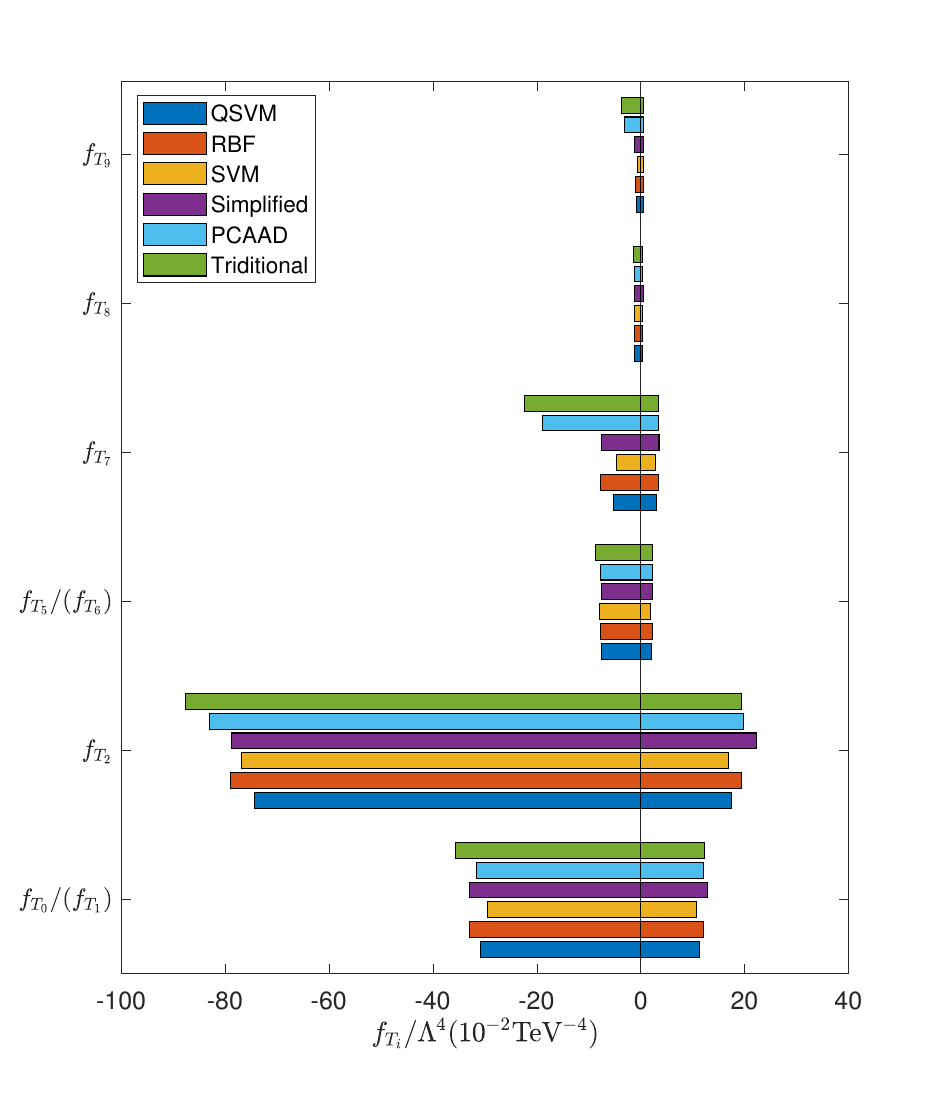}
\includegraphics[width=0.4\textwidth]{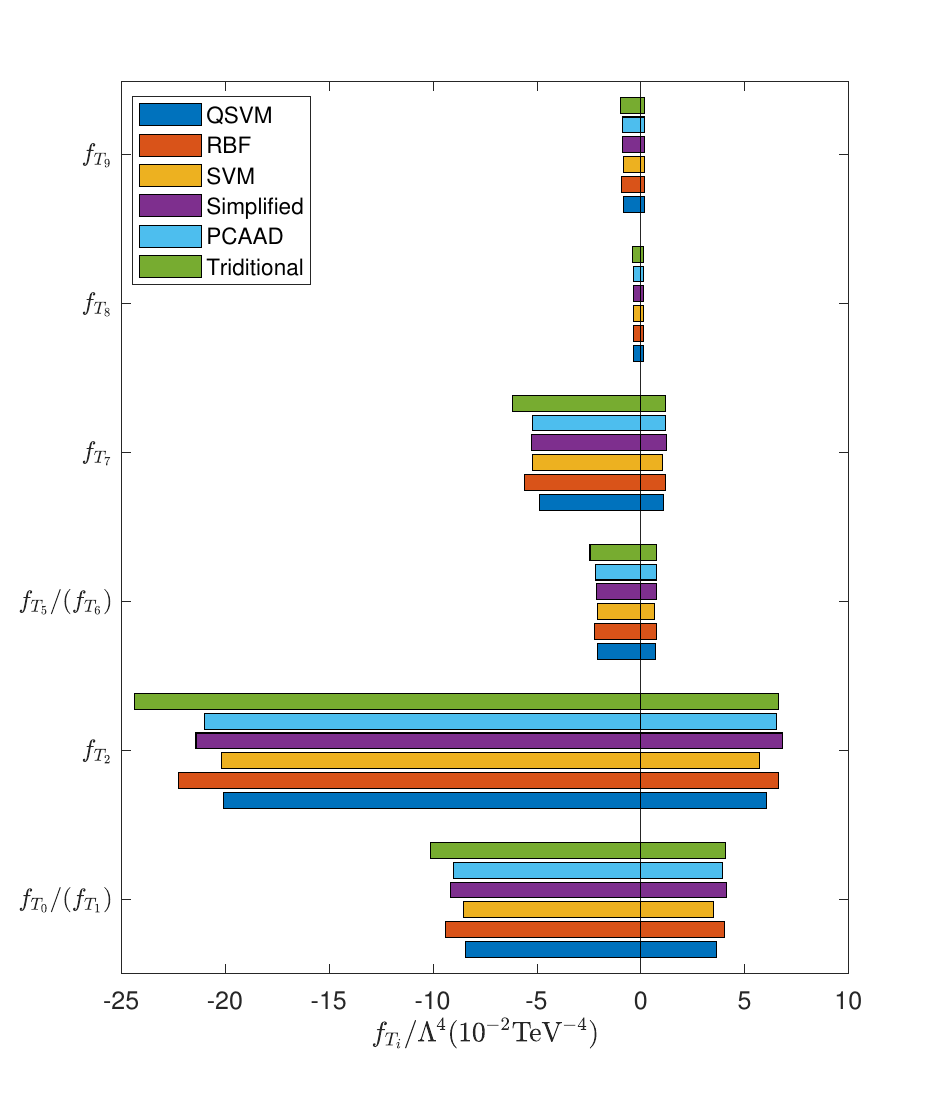}
\includegraphics[width=0.4\textwidth]{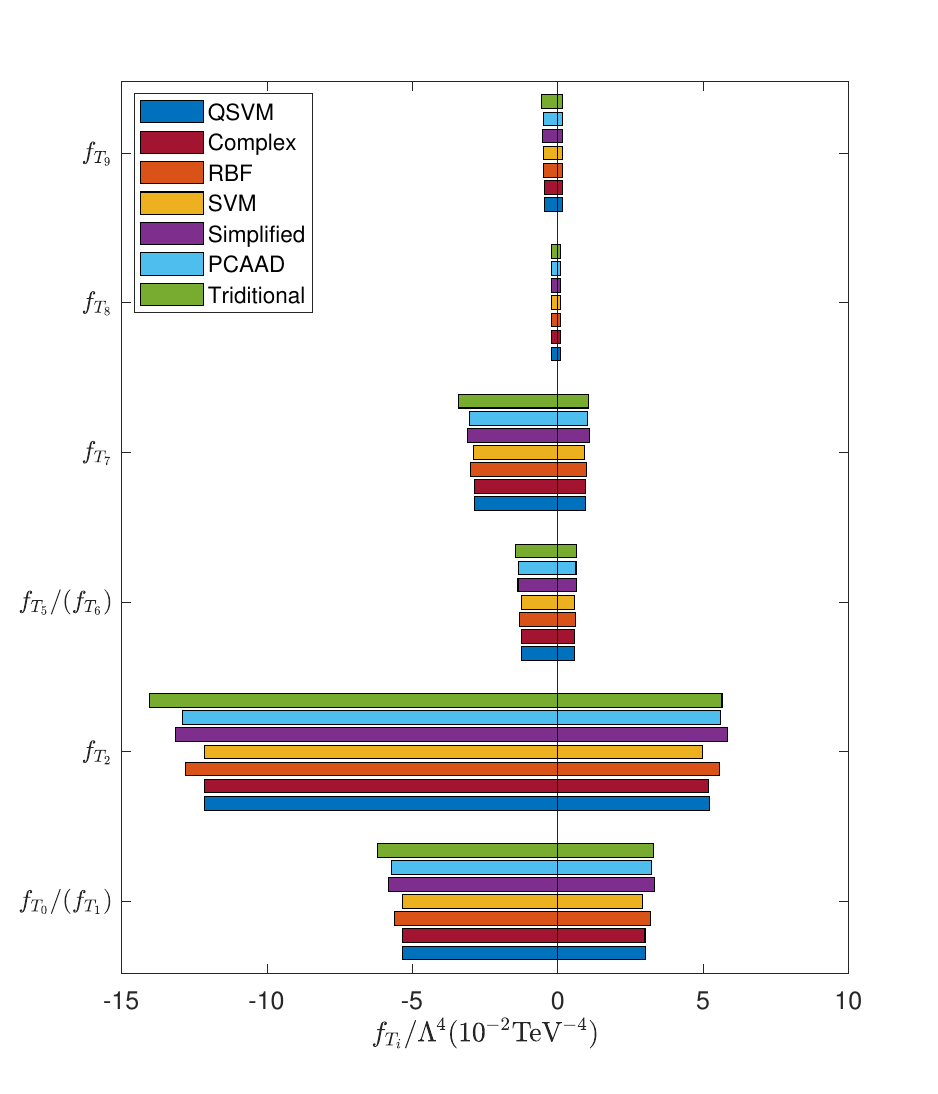}
\caption{\label{fig:compare}
Comparison of the expected coefficient constraints obtained by different algorithms. 
The top-left panel corresponds to $\sqrt{s} = 3\;{\rm TeV}$, the top-right panel corresponds to $\sqrt{s} =10\;{\rm TeV} $, the bottom-left panel corresponds to $\sqrt{s} =14\;{\rm TeV}$, and the bottom-right panel corresponds to $\sqrt{s} =30\;{\rm TeV}$.}
\end{center}
\end{figure*}

This paper uses same data sets as those in Refs~\cite{Yang:2020rjt,Dong:2023nir}, but with different event selection strategies. 
We compare the expected coefficient constraints in this paper with those in Refs.~\cite{Yang:2020rjt,Dong:2023nir}, the results are shown in Fig.~\ref{fig:compare}, which demonstrates the bounds computed by the SVM and QSVM used in this paper are tighter than those in Refs.~\cite{Yang:2020rjt,Dong:2023nir}. 

Note that, Ref.~\cite{Yang:2020rjt} uses traditional event selection strategy, and Ref.~\cite{Dong:2023nir} uses a ML AD algorithm based on principal component analysis.
It can also be shown that, the `simplified' event selection strategy has compatible efficiency.
Therefore, we can conclude that the SVM and QSVM algorithms are useful tools to optimize the event selection strategy to search for NP signals, not to mention that QSVM has the potential to cope with the future development in quantum computing.

\begin{figure}[htpb]
\begin{center}
\includegraphics[width=0.48\hsize]{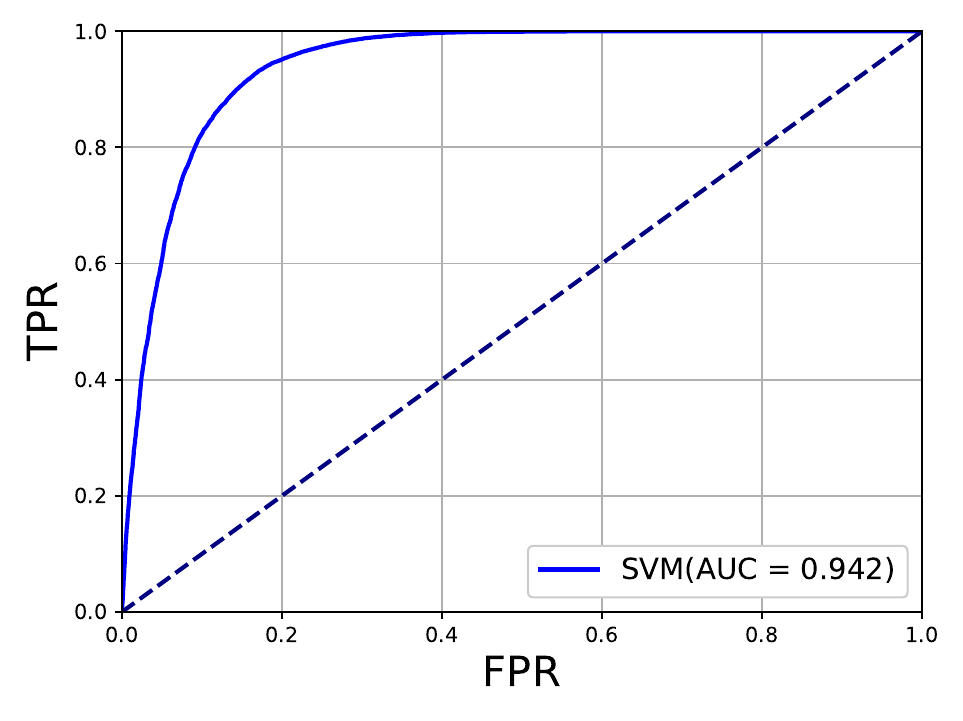}
\includegraphics[width=0.48\hsize]{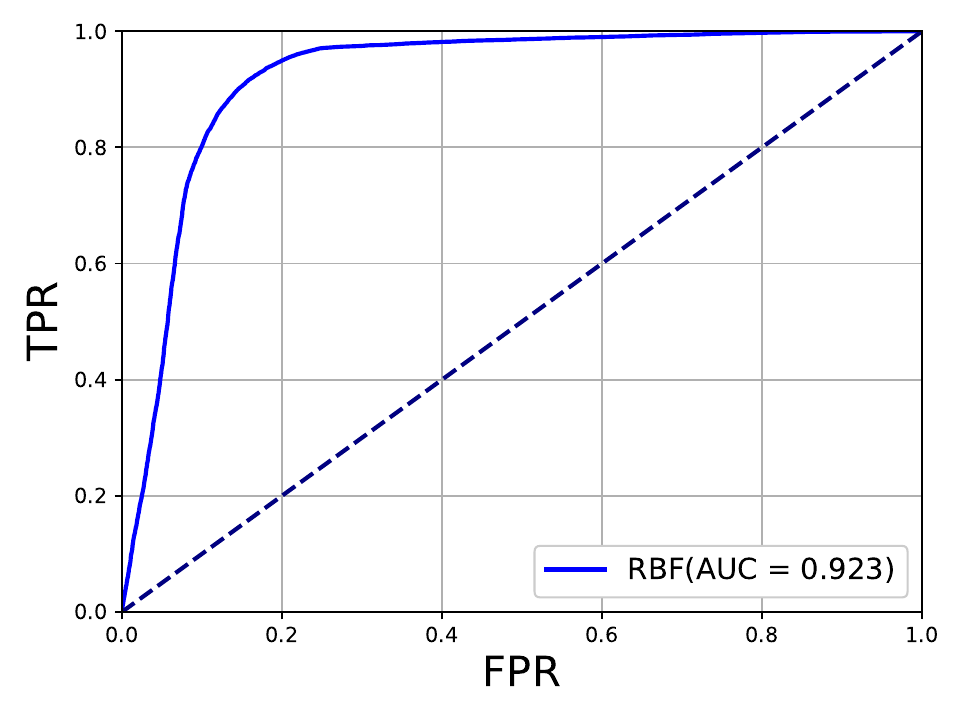}
\includegraphics[width=0.48\hsize]{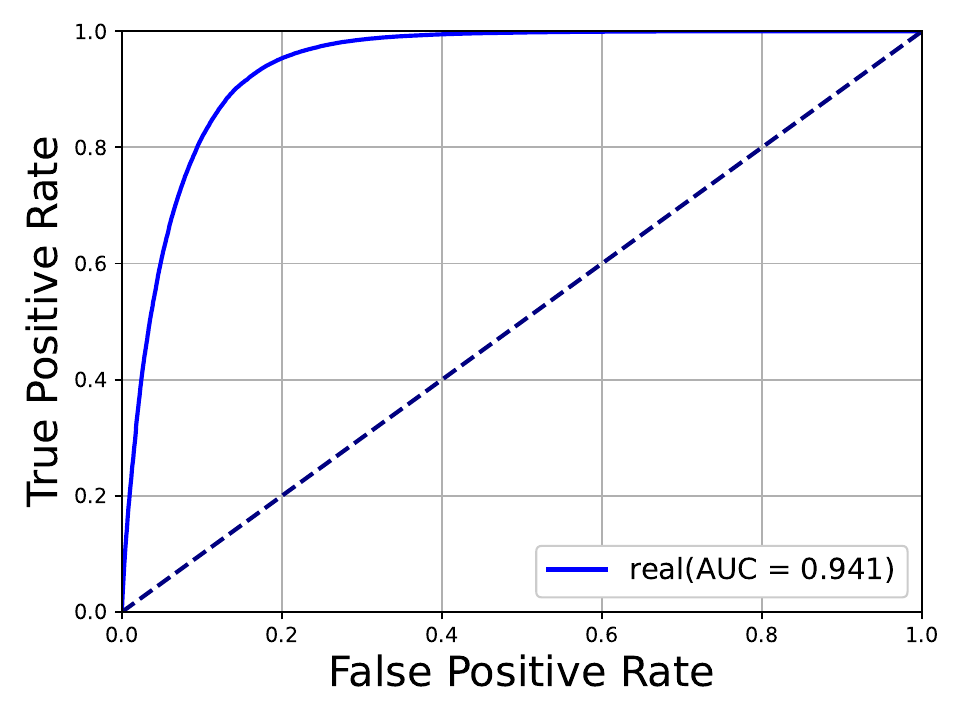}
\includegraphics[width=0.48\hsize]{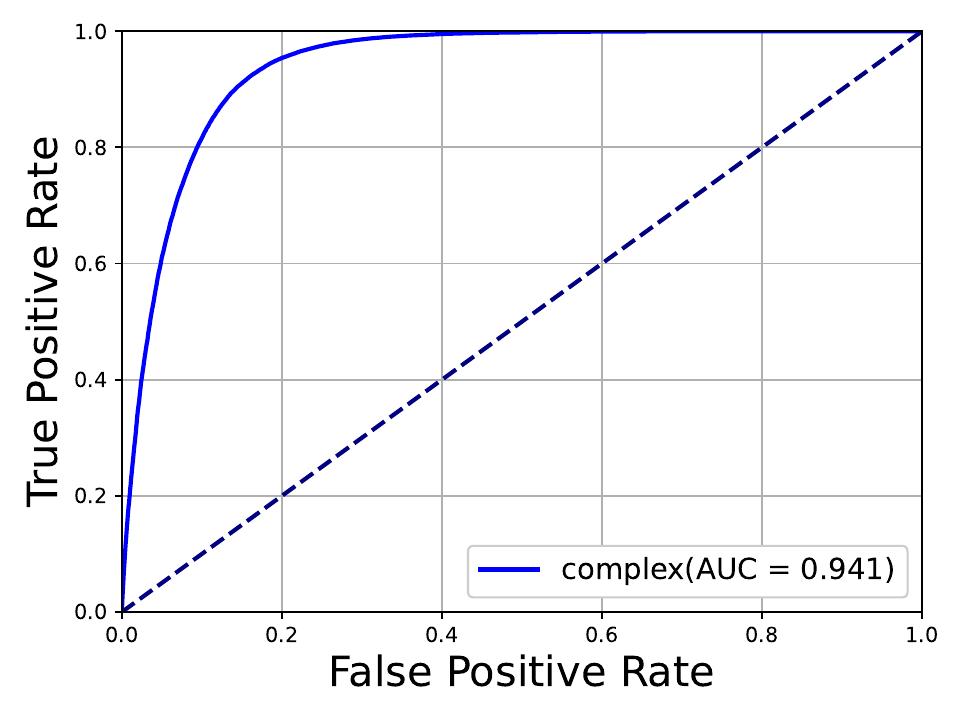}
\caption{\label{fig:aucy}
ROCs of SVMs as well as QSVMs with different kernel functions at $\sqrt{s}=30\;{\rm TeV}$ for events from the SM and $O_{T_0}$ contributions.
The top-left corner corresponds to the linear kernel SVM, the top-right corner corresponds to the RBF kernel, the bottom-left corner corresponds to the QSVM, and the bottom-left corner corresponds to the AUC diagram of the QSVM with complex vectors.}
\end{center}
\end{figure}

Another metric to compare the performance of the classifiers is is receiver operating characteristic~(ROC).
The ROC curve plots the false positive rate~(FPR) on the horizontal axis and the true positive rate~(TPR) on the vertical axis. 
The FPR is the ratio of negative samples incorrectly classified as positive to the total number of negative samples. 
The TPR is the ratio of correctly classified positive samples to the total number of positive samples. 
Ideally, one wants the ROC curve to be as close to the upper left corner as possible. 
The area under the ROC curve~(AUC) quantifies the overall performance, with a higher AUC indicating better prediction accuracy.

As can be seen in Fig.~\ref{fig:aucy}, the linear kernel SVM is the most accurate~($AUC=0.942$), and the accuracies of QSVM and `complex' ~($AUC=0.941$) are almost as same as the one for the linear kernel SVM at $\sqrt{s} = 30\;{\rm TeV}$ for the SM and $O_{T_0}$.
This is consistent with the results in Fig.~\ref{fig:compare}, i.e., the results for QSVM and SVM are of the same order of magnitude, with SVM being slightly better for some operators and c.m. energies, and QSVM being slightly better in other cases.

\section{\label{sec7}Summary}

The search for NP signals involves analyzing a large amount of data, so efficiency becomes a very important consideration in both the phenomenological studies and the experiments.
Both ML algorithms and quantum computers offer the potential to improve the efficiency.
In this paper, a procedure is proposed to use SVM and QSVM to optimize event selection strategies to search for the signals from the NP.

The tri-photon process at a muon collider, which is sensitive to the dimension-8 operators contributing to the aQGCs, is considered as an example.
The optimized event selection strategies are trained with SVM and QSVM.
Then, the expected constraints are calculated, which demonstrate the effectiveness of SVM and QSVM in searching NP signals, therefore SVM and QSVM can be used in AD tasks such as searching NP signals.

This study verifies the feasibility of the QSVM. 
There are also opportunities to further improve the capabilities of QSVM by using a feature space composed of more observables, or by using more complicated quantum kernel functions, or by using optimized algorithms such as multi-state swap test.
Therefore, it can be concluded that QSVM is suitable for phenomenological studies of NP, especially with the help of future developments in quantum computing.

\begin{acknowledgement}
This work was supported in part by the National Natural Science Foundation of China under Grants Nos. 11905093 and 12147214, the Natural Science Foundation of the Liaoning Scientific Committee Nos.~LJKZ0978 and JYTMS20231053, and the Outstanding Research Cultivation Program of Liaoning Normal University (No.21GDL004).
\end{acknowledgement}

\appendix

\section{\label{sec:ap1}The parameters used in the standardization methods}

\begin{table}[htbp]
\centering
\begin{tabular}{c|c|c|c} 
\hline
  $\sqrt{s} ({\rm TeV})$ & $\bar p^{1} ({\rm GeV})$& $ \bar p^{2} ({\rm GeV})$ & $ \bar p^{3} ({\rm GeV}) $ \\ 
  \hline
  $3$ & $1.379\times 10^{3}$&$1.172\times 10^{3}$&$4.481 \times 10^{2}$ \\
  \hline
 $10$ & $4.619\times 10^{3}$& $3.978\times10^{3}$&$1.403\times10^{3}$\\
  \hline
 $14$ &$6.470\times 10^{3}$& $5.584\times 10^{3}$& $1.946\times 10^{3}$\\
  \hline
 $30$ &$1.389\times 10^{4}$&$1.205\times 10^{4}$&$ 4.067\times 10^{3}$\\
  \hline
  $\sqrt{s} ({\rm TeV})$ & $ \bar p^{4} ({\rm GeV})$& $ \bar p^{5} ({\rm GeV})$& $ \bar p^{6}({\rm GeV}) $  \\ 
  \hline
  $3$ &$ 8.815\times 10^{2}$&$7.566\times 10^{2}$ &$3.203\times 10^{2}$\\
  \hline
 $10$  &$2.948\times 10^{3}$&$2.542\times 10^{3}$& $1.017\times 10^{3}$\\
  \hline
 $14$ &$4.152\times 10^{3}$&$3.583\times 10^{3}$& $1.413\times 10^{3}$\\
  \hline
 $30$ &$8.873\times 10^{3}$&$7.674\times 10^{3}$&$2.967\times 10^{3}$\\
  \hline
  $\sqrt{s} ({\rm TeV})$ & $ \bar p^{7}$& $ \bar p^{8} $ & $ \bar p^{9}$\\ 
  \hline
  $3$ & $-4.854\times 10^{-3}$&$3.911\times 10^{-3}$&$6.595\times 10^{-3}$\\
  \hline
 $10$&$-2.743\times 10^{-3}$ & $3.534\times 10^{-3}$ &$4.323\times 10^{-3}$\\
  \hline
 $14$ &$-3.210\times 10^{-3}$& $7.557\times 10^{-3}$&$1.268\times 10^{-2}$\\
  \hline
 $30$ &$-1.055\times 10^{-2}$& $1.058\times 10^{-2}$&$3.782\times 10^{-2}$\\
  \hline
  $\sqrt{s} ({\rm TeV})$ & $ \bar p^{10}$& $\bar p^{11}$ & $ \bar p^{12}$\\ 
  \hline
  $3$ &$3.637$&$3.037$&$1.877$\\
  \hline
 $10$&$3.661 $ & $2.994$ &$1.911$\\
  \hline
 $14$ &$3.655$&$2.978$& $1.916$\\
  \hline
 $30$ &$3.668$& $2.949$&$1.943$\\
  \hline
  $\sqrt{s} ({\rm TeV})$ & $ \bar p^{13}({\rm GeV})$& $ \bar p^{14}({\rm GeV}) $ & $ \bar p^{15} ({\rm GeV}) $ \\ 
  \hline
  $3$ & $2.487\times 10^{3}$&$1.288\times 10^{3}$&$7.307\times 10^{2}$\\
  \hline
 $10$ & $8.391\times 10^{3}$ &$4.009\times 10^{3}$&$2.312\times 10^{3} $\\
  \hline
 $14$ &$1.177\times 10^{4}$& $5.534\times 10^{3}$&$3.205\times 10^{3}$\\
  \hline
 $30$ &$2.533\times 10^{4}$& $1.150\times 10^{4}$&$6.725\times 10^{3}$\\
  \hline
\end{tabular}
\caption{The mean values of $j$-th feature $p^j$
over the SM training data set at 
$\sqrt{s}=3\;{\rm TeV}$,
$10\;{\rm TeV}$, $ 14\;{\rm TeV}$, 
and $ 30\;{\rm TeV}$, respectively.}
\label{table:mean}
\end{table}

\begin{table}[htbp]
\centering
\begin{tabular}{c|c|c|c} 
\hline
  $\sqrt{s} ({\rm TeV})$ & $ d^{1} ({\rm GeV})$& 
  $ d^{2} ({\rm GeV})$ & $ d^{3} ({\rm GeV}) $ \\ 
  \hline
  $3$ & $1.234\times 10^{2}$&$2.107\times 10^{2}$&
  $ 2.932 \times 10^{2}$ \\
  \hline
 $10$&$4.179\times10^{2}$&$7.388\times10^{2}$&
 $1.032\times10^{3}$\\
  \hline
 $14$ &$5.893\times 10^{2}$& $1.044\times 10^{3}$& $1.463\times 10^{3}$\\
  \hline
 $30$ &$1.268\times 10^{3}$&$ 2.263\times 10^{3}$&$ 3.179\times 10^{3}$\\
  \hline
  $\sqrt{s} ({\rm TeV})$ & $ d^{4} ({\rm GeV})$& $ d^{5}({\rm GeV}) $& $ d^{6}({\rm GeV}) $  \\ 
  \hline
  $3$ &$ 3.602\times 10^{2}$&$3.139\times 10^{2}$ &$2.593\times 10^{2}$\\
  \hline
 $10$  &$1.213\times 10^{3}$&$1.061\times 10^{3}$& $8.912\times 10^{2}$\\
  \hline
 $14$ &$1.702\times 10^{3}$&$1.491\times 10^{3}$& $1.261\times 10^{3}$\\
  \hline
 $30$ &$3.647\times 10^{3}$&$3.200\times 10^{3}$&$2.714\times 10^{3}$\\
  \hline
  $\sqrt{s} ({\rm TeV})$ & $ d^{7}$& $ d^{8} $ & $ d^{9}$\\ 
  \hline
  $3$ & $1.238$&$1.208$&$1.148$\\
  \hline
 $10$&$1.242$ & $1.219$ &$1.154$\\
  \hline
 $14$ &$1.235$& $1.123$&$1.157$\\
  \hline
 $30$ &$1.241$& $1.222$&$1.151$\\
  \hline
  $\sqrt{s} ({\rm TeV})$ & $ d^{10}$& $d^{11}$ & $ d^{12}$\\ 
  \hline
  $3$ &$7.922\times 10^{-1}$&$6.937\times 10^{-1}$&$7.313\times 10^{-1}$\\
  \hline
  $10$&$8.130\times 10^{-1} $ & $7.285\times 10^{-1}$ &$7.659\times 10^{-1}$\\
  \hline
 $14$ &$8.156\times 10^{-1}$&$7.376\times 10^{-1}$& $7.754\times 10^{-1}$\\
  \hline
 $30$ &$8.239\times 10^{-1}$& $7.538\times 10^{-1}$&$7.939\times 10^{-1}$\\
  \hline
  $\sqrt{s} ({\rm TeV})$ & $ d^{13}({\rm GeV})$& $ d^{14}({\rm GeV}) $ & $ d^{15} ({\rm GeV}) $ \\ 
  \hline
  $3$ & $3.549\times 10^{2}$&$5.463\times 10^{2}$&$4.434\times 10^{2}$\\
  \hline
 $10$ & $1.239\times 10^{3}$ &$2.062\times 10^{3}$&$1.547\times 10^{3} $\\
  \hline
 $14$ &$1.753\times 10^{3}$& $2.961\times 10^{3}$&$2.197\times 10^{3}$\\
  \hline
 $30$ &$3.803\times 10^{3}$& $6.599\times 10^{3}$&$4.791\times 10^{3}$\\
  \hline
\end{tabular}
\caption{Same as Table~\ref{table:mean} but for the standard deviations.}
\label{table:standarddevition}
\end{table}
In the case of classical SVM methods, a z-score standardization is used, as shown in Eq.~(\ref{eq.zscore}), where $\bar{p}^{j}$ and $d^j$ are the mean values and the standard deviations of the $j$-th features over the SM training data sets at different c.m. energies, and listed in Tables~\ref{table:mean} and \ref{table:standarddevition}.

\begin{table}[htbp]
\centering
\begin{tabular}{c|c|c|c} 
\hline
 $\sqrt{s}({\rm TeV})$ & $ p^{1}_{\rm \rm min}({\rm GeV})$& $ p^{2}_{\rm min} ({\rm GeV})$ & $ p^{3}_{\rm min}({\rm GeV}) $ \\ 
 \hline
 $3$ & $ 1.012 \times 10^{3} $ & $7.599 \times 10^{2}$&$1.004\times 10^{1}$\\
 \hline
 $10$ &$3.343\times 10^{3}$&$2.509\times 10^{3}$&$1.089\times 10^{1}$\\
  \hline
 $14$ &$4.750\times 10^{3}$& $3.536\times 10^{3}$& $9.773$\\
  \hline
 $30$ &$1.008\times 10^{4}$&$7.594\times 10^{3}$&$9.598$\\
  \hline
  $\sqrt{s} ({\rm TeV})$ & $p^{4}_{\rm min} ({\rm GeV})$& $ p^{5}_{\rm min}({\rm GeV}) $ & $ p^{6}_{\rm min}({\rm GeV})$  \\ 
  \hline
  $3$ & $ 1.831\times 10^{2}$&$1.413\times 10^{2}$&$ 9.540$\\
 \hline
 $10$ & $6.924\times 10^{2}$&$4.470\times 10^{2}$&$9.600$\\
  \hline
 $14$ &$8.161\times 10^{2}$&$7.068\times 10^{2}$&$9.550$\\
  \hline
 $30$ &$1.827\times 10^{3}$&$1.387\times 10^{3}$&$9.450$\\
  \hline
  $\sqrt{s} ({\rm TeV})$ & $ p^{7}_{\rm min}$& $p^{8}_{\rm min} $ & $ p^{9}_{\rm min}$ \\ 
  \hline
  $3$ &$-2.496$&$-2.492$&$-2.498$\\
 \hline
 $10$ & $-2.497$&$-2.498$&$-2.489$\\
  \hline
 $14$ &$-2.486$&$-2.496$&$-2.500$ \\
  \hline
 $30$ &$-2.500$&$-2.499$&$-2.497$\\
 \hline
 $\sqrt{s}({\rm TeV})$ & $ p^{10}_{\rm min}$& $ p^{11}_{\rm min} $ & $ p^{12}_{\rm min}$ \\
 \hline
 $3$ &$2.208$&$4.751\times 10^{-1}$&$3.994\times 10^{-1}$\\
 \hline
 $10$ & $2.195$&$4.496\times 10^{-1}$&$4.049\times 10^{-1}$\\
  \hline
 $14$ &$2.255$&$4.154\times 10^{-1}$&$4.083\times 10^{-1}$\\
  \hline
 $30$ &$2.154$&$4.123\times 10^{-1}$&$4.093\times 10^{-1}$\\
 \hline
  $\sqrt{s}({\rm TeV})$ & $ p^{13}_{\rm min}({\rm GeV})$& $ p^{14}_{\rm min} ({\rm GeV})$ & $ p^{15}_{\rm min}({\rm GeV}) $ \\
  \hline
 $3$ &$1.760\times 10^{3}$&$6.684\times 10^{1}$&$2.530\times 10^{1}$\\
 \hline
 $10$ &$5.804\times 10^{3}$&$7.278\times 10^{1}$&$6.008\times 10^{1}$\\
  \hline
 $14$ &$ 8.203\times 10^{3}$ &$7.279\times 10^{1}$&$8.577\times 10^{1}$\\
  \hline
 $30$ &$1.746\times 10^{4}$&$ 1.452\times 10^{2}$&$ 1.011\times 10^{2}$\\
 \hline
\end{tabular}
\caption{The minimal values of $p_{i}$ over the SM training set contains $1000$ events at different c.m. energies.}
\label{table:min}
\end{table}

\begin{table}[htbp]
\centering
\begin{tabular}{c|c|c|c} 
\hline
 $\sqrt{s}({\rm TeV})$ & $ p^{1}_{\rm max}({\rm GeV})$& $ p^{2}_{\rm max} ({\rm GeV})$ & $ p^{3}_{\rm max}({\rm GeV}) $\\ 
 \hline
 $3$ & $ 1.534 \times 10^{3} $ & $1.510 \times 10^{3}$&$9.852\times 10^{2}$\\
 \hline
 $10$ &$5.132\times 10^{3}$&$5.069\times 10^{3}$&$3.350\times 10^{3}$\\
  \hline
 $14$ &$7.198\times 10^{3}$& $7.090\times 10^{3}$& $4.639\times 10^{3}$\\
  \hline
 $30$ &$1.543\times 10^{4}$&$1.522\times 10^{4}$&$9.919\times 10^{3}$\\
  \hline
  $\sqrt{s} ({\rm TeV})$ & $p^{4}_{\rm max} ({\rm GeV})$& $ p^{5}_{\rm max}({\rm GeV}) $ & $ p^{6}_{\rm max}({\rm GeV})$  \\ 
  \hline
  $3$ & $1.530\times 10^{3}$& $1.496\times 10^{3}$& $ 9.794\times 10^{2}$\\
 \hline
 $10$ & $5.026\times 10^{3}$&$4.999\times 10^{3}$& $3.273\times 10^{3}$\\
  \hline
 $14$ &$7.080\times 10^{3}$&$6.955\times 10^{3}$&$4.542\times 10^{3}$\\
  \hline
 $30$ & $1.513\times 10^{4}$&$1.495\times 10^{4}$&$9.789\times 10^{3}$\\
  \hline
  $\sqrt{s} ({\rm TeV})$ & $ p^{7}_{\rm max}$& $p^{8}_{\rm max} $ & $ p^{9}_{\rm max} $ \\ 
  \hline
  $3$&$2.499$&$2.499$&$2.494$\\
 \hline
 $10$ &$2.488$&$2.491$& $2.499$\\
  \hline
 $14$ &$2.484$&$2.500$&$2.498$\\
  \hline
 $30$ &$2.497$&$2.492$&
  $2.495$\\
 \hline
  $\sqrt{s} $ & $p^{10}_{\rm max}$& $ p^{11}_{\rm max} $ & $ p^{12}_{\rm max}$  \\ 
  \hline
  $3$ &$5.850$&$5.447$&$4.168$\\
 \hline
 $10$ &$5.875$& $5.477$&$4.689$\\
  \hline
 $14$ & $5.865$&$5.520$&$5.098$\\
  \hline
 $30$ &$5.896$&$5.449$&$5.119$\\
  \hline
  $\sqrt{s} ({\rm TeV})$ & $ p^{13}_{\rm max} ({\rm GeV})$& $p^{14}_{\rm max}  ({\rm GeV})$ & $ p^{15}_{\rm max} ({\rm GeV})$ \\ 
  \hline
  $3$ &$3.033\times10^{3}$&$2.130\times10^{3}$&$1.728\times 10^{3}$\\
 \hline
 $10$ &$1.015\times 10^{4}$&$7.002\times 10^{3}$&$5.746\times 10^{3}$\\
  \hline
 $14$ &$ 1.421\times 10^{4}$ &$9.863\times 10^{3}$&$7.910\times 10^{3}$\\
  \hline
 $30$ &$3.051\times 10^{4}$&$ 2.110\times 10^{4}$&$ 1.727\times 10^{4}$\\
 \hline
\end{tabular}
\caption{Same as Table~\ref{table:min} but for the maximal values of $p_{i}$.}
\label{table:max}
\end{table}
In the case of quantum kernel SVM methods, a min-max normalization is used, as shown in Eq.~(\ref{eq.x}). The $p_{i,min}$ and $p_{i,max}$ are the minimal and maximal values of the $i$-th feature over the SM training set, and are summarized in Tables~\ref{table:min} and \ref{table:max}, respectively.

\bibliography{svm}
\bibliographystyle{elsarticle-num}

\end{document}